\documentclass[acmtog,screen]{acmart}

\usepackage{multirow}
\usepackage{color, colortbl}
\definecolor{LightBlue}{rgb}{0.67,0.84,0.9}
\definecolor{LightYellow}{rgb}{0.9,0.9,0.7}
\definecolor{LightRed}{rgb}{0.9,0.67,0.67}
\definecolor{LightGray}{rgb}{0.9,0.9,0.9}
\definecolor{DarkGray}{rgb}{0.66, 0.66, 0.66}
\definecolor{LightGray}{rgb}{0.9, 0.9, 0.9}

\usepackage{tabularx}
\usepackage{longtable}
\usepackage[title]{appendix}
\usepackage{acmart-taps}

\usepackage{caption}
\usepackage{subcaption}
\usepackage{graphicx}
\usepackage{acmart-taps}

\AtBeginDocument{%
  }

\copyrightyear{2023}
\acmYear{2023}
\setcopyright{rightsretained}
\acmConference[FAccT '23]{2023 ACM Conference on Fairness, Accountability, and Transparency}{June 12--15, 2023}{Chicago, IL, USA}
\acmBooktitle{2023 ACM Conference on Fairness, Accountability, and Transparency (FAccT '23), June 12--15, 2023, Chicago, IL, USA}
\acmDOI{10.1145/3593013.3594016}
\acmISBN{979-8-4007-0192-4/23/06}

\begin{document}
\newcommand{\pquote}[1]{``\textit{#1}''}

\title{AI’s Regimes of Representation: A Community-centered Study of Text-to-Image Models in South Asia  
}

\author{Rida Qadri}
\affiliation{
  \institution{Google Research}
  \country{San Francisco, California, USA}
}

\author{Renee Shelby}
\affiliation{
  \institution{Google Research}
  \country{San Francisco, California, USA}
}

\author{Cynthia L. Bennett}
\affiliation{
  \institution{Google Research}
  \country{New York, New York, USA}
}

\author{Remi Denton}
\affiliation{
  \institution{Google Research}
  \country{New York, New York, USA}
}

\renewcommand{\shortauthors}{Qadri et al.}

\begin{abstract}
This paper presents a community-centered study of cultural limitations of text-to-image (T2I) models in the South Asian context. We theorize these failures using scholarship on dominant media regimes of representations and locate them within participants’ reporting of their existing social marginalizations. We thus show how generative AI can reproduce an outsiders gaze for viewing South Asian cultures, shaped by global and regional power inequities. By centering communities as experts and soliciting their perspectives on T2I limitations, our study adds rich nuance into existing evaluative frameworks and deepens our understanding of the culturally-specific ways AI technologies can fail in non-Western and Global South settings. We distill lessons for responsible development of T2I models, recommending concrete pathways forward that can allow for recognition of structural inequalities.
\end{abstract}



\maketitle


\section{Introduction}  
\label{sec:intro}
Emerging FAccT scholarship points to the need for reevaluating the field's dominant  methods and frameworks  for understanding and evaluating AI harms. For instance, there are growing calls for more community-centered work ~\cite{kapania2022} and a re-orientation towards non-Western frameworks of fairness ~\cite{kak2020, amrute2022primer, png2022, sambasivan2021, sambasivan2022, sambasivan2022, mohamed2020decolonial}. However, empirical studies collaboratively investigating AI harms with diverse, global communities are less common, continuing the disconnect between dominant evaluation approaches and the lived experiences of impacted communities ~\cite{birhane2022forgotten}. 
  
In response, we conducted a community-centered study of cultural limitations of text-to-image (T2I) models in the South Asian context, with 36 participants from Pakistan (\textit{n} = 15), India (\textit{n} = 13), and Bangladesh (\textit{n} = 8). Through two-part focus groups, participants co-designed T2I prompts and collectively reflected on model outputs. This study design offered participants agency to articulate their own understandings of model limitations, failures, and potential impacts, drawing from their local cultural knowledge and situated experiences with South Asian representations. Participant conversation and reflections foreground three broad failure modes: failing to generate cultural subjects, amplifying hegemonic cultural defaults, and perpetuating cultural tropes. We contextualize these failures within literature on media and cultural studies, that study the power of media's \textit{regimes of representations} ~\cite[p. 234]{hall1997representation}: controlling narratives and discourses about particular social groups.  Our study shows how cultural limitations of T2I models can participate in and scale such existing harmful media regimes of representation and amplify experiences of socio-cultural marginalization. While the T2I failure modes foregrounded by participants are not necessarily unique to South Asia, their specific articulation in South Asia is contextual and socially situated in the global and regional power dynamics that shape the region.

It is important to note, however, this study is not a systematic evaluation of T2I model capabilities. As such, we do not position our findings as definitive commentary on any one or all T2I models. Nonetheless, our study, albeit a limited form of community engagement, encourages important methodological reflection on common machine learning evaluative and benchmarking practices. Emerging empirical work on T2I evaluations focuses on quantitative evaluation ~\cite{Ge_undated-sx, cho2022dall}, often with bias metrics pre-determined by researchers and practitioners  ~\cite{bianchi2022easily}. While such approaches enable evaluating models at “scale,” without community input they risk centering un-nuanced notions of harm that do not fully account for on-the-ground community experiences in different cultural contexts.

Our work contributes to multiple strands of responsible AI research. First, we respond to prior calls to re-orient algorithmic fairness outside Western contexts ~\cite{sambasivan2020non, sambasivan2021re} by offering the first empirical study of T2I performance in South Asia. Second, by soliciting community perspectives on T2I limitations, we identify novel T2I failure modes and connect model limitations to communities’ lived experiences, expanding the field’s understanding of cultural harms. Third, by historicizing emerging generative image technologies within scholarship on the politics of representation, we identify particular regimes of representation within T2I models, revealing how these regimes draw from and perpetuate marginalizing discourses. Thus, our study offers an example of how qualitative, community-centered research strengthens responsible AI practice through centering local knowledge and expertise.  

\section{Theoretical Background}
\label{sec:background}
\subsection{Text-to-Image Models and AI Harms}
Text-to-image (T2I) generative models allow users to create photorealistic images from free-form and open-ended text prompts ~\cite{radford2021, ramesh2021, saharia2022photorealistic, yu2022scaling}, typically relying on web-scale datasets. Such datasets have been shown to reflect social stereotypes, inequalities, and hierarchies ~\cite{paullada2021data, birhane2021multimodal, birhane2021pyrrhic}, raising concerns about T2I models similarly fostering representational and cultural harms ~\cite{shelby2022sociotechnical, tomasev2022manifestations, Srinivasan2021-zd}. However, unlike other generative AI, such as language ~\cite{bender2021dangers, Weidinger2022} or image caption models ~\cite{wang2022}, computing researchers have yet to articulate the broader landscape of potential harms for generative image models. While empirical research on T2I models is still nascent, studies suggest they can reinforce social hierarchies and replicate dominant stereotypes along axes of gender, skin tone, and culture ~\cite{bianchi2022easily, cho2022dall, wolfe2022contrastive, bansal2022well}. Our work complements and extends this work, by offering the first empirical study of T2I models that centers and engages non-Western communities.

\subsection{Non-Western and Community-Centered Fairness}
There is a growing body of scholarship calling attention to the dominance of Western perspectives and experiences embedded within responsible AI frameworks ~\cite{png2022, kak2020, sambasivan2022, sambasivan2021, sambasivan2020non, prabhakaran2022cultural}, which are not transferable across cultural contexts ~\cite{barabas2020studying, weinberg2022rethinking}. The non-portability of Western frameworks can lead to flawed data and model assumptions, evaluation methods that overlook culturally-specific axes of discrimination, and cultural incongruencies ~\cite{sambasivan2020non, sambasivan2021re, prabhakaran2022cultural}. When operationalized in model testing and evaluation, exclusive use of Western-oriented frameworks risks development of applications that dispossess the identity of non-Western communities ~\cite{mohamed2020decolonial}, by centralizing the epistemologies used and power to build algorithmic systems in the hands of a global minority ~\cite{hanna2020}. Compared to other AI harms, such as \textit{representational} or \textit{allocative harms}, much less attention has been devoted in computing literature to understanding cultural harms, leaving these “under articulated” in the field  ~\cite[p. 18]{shelby2022sociotechnical}. Current approaches to understanding cultural harms focuses on how they can foreclose ways of understanding the social world ~\cite{sadowski2014creating}, leading to systemic erasure ~\cite{devos2022}, proliferating false ideas about cultural groups ~\cite{sambasivan2021}, and exporting Western ideas to the Global South ~\cite{mohamed2020decolonial}. However, the nuanced ways these take shape for different non-Western communities are not well-understood.

More globally inclusive and community-centered approaches to AI fairness and cultural harms require recontextualizing data and model evaluation — with an explicit incorporation of contextual axes of discrimination ~\cite{sambasivan2020non}. Particularly there are calls to meaningfully center different global communities and institutions in knowledge production processes ~\cite{selbst2019fairness, arun2019ai} and incorporate participatory practices that allow production of ML frameworks by impacted communities.  ~\cite{sambasivan2020non}. Combined with community-centered research, ML practices that center reciprocity, reflexivity, and empowerment can help reshift power dynamics between technologists and marginalized communities ~\cite{birhane2022power, kalluri2020}.

\subsection{Regimes of South Asian Representation}
\textit{Representation}, such as through visual media, is the process of creating and communicating meaning about the social world ~\cite{hall1997representation}. There are no “true” representations; rather representations are “historically determined [social] construction(s)…mediated by social, ideological, and cultural processes” ~\cite[p. 115]{desai2000imaging}. The power to represent communities in ways that shape how they are understood can be understood as a \textit{regime of representation} ~\cite{hall1997representation}, a dominant system of media discourse, symbols and images that create particular narratives about already marginalized groups. Hall shows how these representations are not just one off instances but are part of a broader ``regime of power", that is upheld across media systems, controlling and shaping how others see specific groups.   In the Asian context, a dominant regime of representation is \textit{Orientalism}, which refers to a broader system of thought, way of writing, and studying the “Orient,” or Eastern world, that emerged in the 19th century ~\cite{said1978}. As framed by Western geopolitical forces, the “Orient” became a singular stand-in for the numerous cultural and national boundaries of the Asia continent ~\cite{breckenridge1993, inden2001}. Through Orientalism’s \textit{outsider’s gaze}, the West imposed demeaning cultural stereotypes onto Asia: backwards, silently different, passive, and sexualized ~\cite{breckenridge1993}. Creating harmful representations of Asia was important for “dividing up the difference between (Europe, the West, ‘us’) and the strange (the Orient, the East, ‘them’)” ~\cite[p. 43]{said1978}. Thus, Orientalism was not simply a way of thinking about South Asia, but a means to conceptualize the geography of the colonial world that made Asia susceptible to certain kinds of control and geopolitical management ~\cite{breckenridge1993}. 

Orientalist tropes continue in contemporary media about South Asia. These including essentialist representations of the sub-continent as diseased and mentally ill ~\cite{burr2002cultural, doran2016popular}, impoverished ~\cite{brouilette2011, desai2011}, and economically dysfunctional ~\cite{bhagavan2001mis}. Reductive depictions of Asian women as sexually available and exotic ~\cite{maira2008belly, lewis2004rethinking}, or lacking agency through Western understandings of veiling ~\cite{lewis2004rethinking, sonbol1993} are also common. While the range of Orientalist representations vary, they are united in distorting the meaning of a cultural practice or symbol through reduction and simplification ~\cite{nacos2004framing}. 

Regimes of representation are important sites of analysis because they shape hegemonic ways of seeing and knowing about a culture or community, both externally and internally ~\cite{lau2014, lau2009}. Moreover, the reductive stereotypes, miscategorizations, and forms of erasure ~\cite{lau2014} can “block the capacity of marginalized groups … to imagine, describe, and invent [themselves] in ways that are liberatory” ~\cite[p. 2]{hooks1992}. To date, little is known about what regimes of representation T2I models contain and perpetuate, particularly as defined by South Asian communities. Recognizing these regimes is necessary to disrupt their harmful impacts.  

\section{Methodology}
\label{sec:methods}
In alignment with broader calls for developing non-Western and community-centered responsible AI practices ~\cite{mohamed2020decolonial, sambasivan2021}, we engaged participants from three South Asian countries through focus groups and a survey to: (1) collaboratively develop culturally-specific text prompts, (2) collectively reflect upon images generated by T2I models in response to culturally-specific text prompts, and (3) understand participant experiences of generated imagery. In this section, we provide context and details about our methodology. For additional details, please see Appendix \ref{sec:supplementary_methods}.

\subsection{Site of Study: South Asia}
We focus this study on South Asia, reflecting the cultural expertise of the lead author. As our goal is to localize understandings of T2I cultural failure modes, we recruited from three different South Asian nation-states: Bangladesh, India, and Pakistan. We recognize South Asia is a rich and complex region with many diverse cultures that could be subdivided along multiple other axes (e.g., gender, religion) ~\cite{bose2017}. The chosen nation-states share cultural histories with enough common overlap to facilitate a cohesive yet nuanced analysis and allow for more analytical comparisons.

\subsection{Study Participants} 
We recruited a purposive sample \cite{palinkas2015purposeful} of participants with cultural knowledge of any one of the three nation-states in the form of lived experience, professional affiliation, and/or academic study about Pakistan, India, or Bangladesh. Purposive sampling was used to ensure diversity across the source of cultural knowledge and nation-state. We asked prospective participants to self-identify with one of the nation-states and describe their experience with the selected nation-state; we did not exclude participants based on citizenship or current residency. This allowed us to capture prospective participants with deep cultural knowledge in the South Asian diaspora who are living outside their home countries. Other inclusion criteria required participants have English-language proficiency and be at least 18 years old.

We recruited through targeted emails to (1) academic listservs focused on South Asian Studies programs in North America, Europe, and Asia registered with the Association for Asian Studies, as well as computing research listservs to recruit potential participants with diverse domain expertise directly relevant to our research aims; (2) cultural institutions in Pakistan, India, and Bangladesh; and (3) through the research team's professional networks in South Asia. We received 219 responses. We excluded those who had only visited a country for tourism, currently work for a major technology firm, and “spam” replies with misidentified provinces or languages.  We invited 52 eligible participants and; 36 people ultimately participated (Pakistan (\textit{n} = 15); India (\textit{n} = 13); Bangladesh (\textit{n} = 8). While we did not systematically recruit for intra-national cultural knowledge (e.g., linguistic and regional diversity), our sample covered ten linguistic groups and fourteen sub-national regional groups within South Asia. We had participant diversity across occupational expertise, including 17 academic researchers, 9 “cultural workers” employed in cultural industries, such as museum curation and the arts, and 10 participants with lived experience of the cultural contexts we were studying, but not necessarily professional experience in cultural industries. All participants received a localized equivalent of \$300 USD in thanks for their participation. 

\subsection{Method of Engagement: Focus Groups}
We crafted a study design that facilitates collective engagement and conversation: focus groups. For studying culture, focus groups offer an effective means of accessing culturally-specific knowledge ~\cite{rodriguez2011culturally, colucci2008use} and can “facilitate culturally sensitive research” ~\cite[p. 777]{Hughes2002using}, as the setting helps foster cohesiveness among participants ~\cite{krueger2000practical}, where interactions among participants generate important information ~\cite{Vaughn2012-ij} on cultural representation and cultural harms ~\cite{onwuegbuzie2009qualitative}. We also created opportunities for anonymous feedback through digital whiteboards.

Each participant attended two 90-minute focus groups composed of between 7 and 9 participants from the same country to facilitate rapport ~\cite{kirk1986reliability} and allow participants to focus on the histories and cultures most relevant to them. The first focus group was structured around discussion questions and interactive activities to understand how participants defined “good” and “bad” cultural representations they had encountered in different media and how participants might assess “good” and “bad” representations in AI-generated imagery. To orient participants around the capabilities of T2I models we shared sample generated images of Western and South Asian cultural subject matter. The cross-cultural images served as a point of comparison in participant reflections. Following the first focus group, participants completed a survey submitting full-sentence text prompts and suggesting up to five examples of cultural events, landmarks, art styles and/or artists, historic events, figures, and characters they felt would enable the assessment of T2I models. 

During the second focus group, participants reviewed generated images from 4-5 prompts, seeing four images per prompt. While this number of images is not sufficient to draw inferences about statistical distributions in T2I model outputs, the smaller sample enabled participants to conduct deeper reflections on generated images in alignment with study goals. During these reflections, we requested participants specifically identify what they thought generated images ``got right’’ and what models did “poorly.” Following individual reflection, we facilitated discussion on the possibilities and risks of T2I models. We deliberately kept discussion questions open–ended to give participants agency to focus on what they found most important.

\subsection{Developing Prompts and Generating Images}
Between the first and second focus group, the research team synthesized participant's prompt suggestions, tested various prompts, and generated images using four state-of-the-art T2I models \cite{stable-diffusion, parti, imagen, dalle}. We constructed prompts based on participant suggestions with the aim of increasing quality and coverage of cultural references in the study.  As the study’s focus is on cultural limitations, we utilized prompts that minimized the likelihood of non-cultural failures to ensure we made best use of participant’s time and expertise.  For example, a general model failure for the prompt, “A day in Lahore,” might result in images of daylight, rather than the city; however, rewording the prompt as “People spending their day in Lahore” led to images reflecting the model’s learned associations with daily life in Lahore.

The final corpus included 120 prompts. We randomly assigned each prompt to two of four state-of-the-art T2I models for image generation. We then selected the first two images from each model for participant feedback in the second focus group. Each participant was assigned to review the selected images from 4-5 text prompts. This scoping allowed for model coverage in the study, while keeping the number of image outputs manageable for participants to comment on during the focus group.

\subsection{Data Analysis}
All focus groups were video-recorded, transcribed, and thematically analyzed ~\cite{Braun_Clarke_2006, Braun_Clarke_2021}. We also compiled the video conference chat and participants’ written feedback into the data corpus.
All four authors participated in data analysis, which involved iterative and collaborative development and discussion of codes and themes, drawing upon the reflective thematic analysis approach described in ~\cite{Braun_Clarke_2006, Braun_Clarke_2021}.
After reviewing and developing initial codes for all the focus group data, the research team shared, arbitrated, and iterated on codes, developing preliminary themes during a series of data sessions ~\cite{braun2012thematic}, aligning on themes that cut across the focus groups while attending to poignant differences between the cultural centers of focus. 
Interpreting raw data with relevant cultural scholarship textured participant contributions ~\cite{timmermans2012theory} with broader scale explanations and implications of cultural phenomena that participants could only briefly reference at the focus groups given time constraints. We introduce participant quotes with alphanumeric identifiers, providing their country to better contextualize participant comments. Some quotes were provided anonymously during interactive exercises, and they will only have the identifier "A" after the country-group.

\subsection{Limitations}
While our work offers important insights on cultural failure modes of T2I models, our methodology has limitations. A purposive sampling strategy and focus on three nation-states within a diverse region does not necessarily lead to insights that generalize across South Asia, especially as we did not systematically recruit for participant diversity, such as caste, ethno-linguistic identity, and class. Having English language proficiency as inclusion criteria, and participant and researcher academic affiliations also suggest participants may largely reflect upper-class and urban-dwelling subsets of South Asia. Finally, for time and safety reasons, we did not generate images with participants in real-time; and thus, researchers remained arbiters of shown images. In this way, we characterize this study as community-centered rather than participatory. Our process of prompt synthesis and image generation did not systematically disentangle the effects of different words or phrases on generated images, limiting our ability to draw inferences regarding why certain representations were emerging. We leave this important research direction to future work. 

\section{Findings}
\label{sec:findings}
\subsection{Regimes of Representation in T2I Models}
\label{sec:limitations}
In this section, we outline three model failures foregrounded by South Asian participants and how they characterized these failures as reinscribing dominant \textit{regimes of representation} ~\cite{hall1997representation}; in this case, the hegemonic ways of seeing, knowing and representing South Asia they experience. Broadly, participants were interested in both the accuracy of cultural subject matter recognition and nuances of cultural representations in T2I generated imagery. As P35, from Pakistan, summarized \textit{“[if I put in a particular figure, historical event, or allegory], does [the model] get what I'm trying to say, first of all? Is there a kind of understanding or legibility? But then within that, what kind of visual representation do you get? Do you get a kind of Orientalist, portraiture rendering? Do you get an image that looks closer to maybe South Asian renderings?”} 

Drawing from this call for cultural recognition, we present three failure modes that encapsulate participant concerns about model accuracy and representations: (1) failing to recognize cultural subjects: generated imagery fails to depict a culture’s subject matter; (2) amplifying cultural defaults: culture’s subject matter in generated images defaults to particular hegemonic cultures; and (3) perpetuating cultural tropes: generated images contain stereotypes and tropes associated with particular cultures.

\subsubsection{Failing to Recognize Cultural Subjects}  
Participants shared their desire to test T2I models' ability to generate cultural artifacts, history, and practices from South Asian cultures. Importantly, participants were not looking for absolute accuracy in each image, and emphasized its impossibility for topics with multiple realities and possible renderings (e.g., a South Asian family). Rather, they adjudicated accuracy based on whether the cultural subject matter had a canonical rendering (e.g., historical figures like Indira Gandhi and architectural landmarks like Badshahi Mosque), or essential canonical elements (e.g., the correct sporting equipment for cricket scenes, the proper landscape for a region, or the art style of Sadequain). Participants emphasized their concerns about accuracy extended to equitable cross-cultural performance, as P17 from India explained: \pquote{if [the machine] can recognize the style of Picasso, then, you know, is it equally possible for a machine to recognize the style of Warli paintings [a style of Indian folk art]?} Through their reviews of generated imagery, participants identified different dimensions of "failure to recognize cultural subjects," from total failure to partial legibility lacking cultural specificity. 

Across all countries, South Asian participants identified examples where  models completely failed to depict important cultural subject matter specified in text prompts. For instance, models totally failed to render the styles of famous artists from India (e.g., Tagore), Pakistan (e.g., Gulgee, Sadequain), and Bangladesh (e.g., Zainul Abdein). Participants described how such total failures were particularly frustrating as generated images shown during the first focus group reflected the painting styles of Monet, Picasso, and Rembrandt in easily recognizable ways. These cross-cultural failures were obvious, as P18 from India reflected: \textit{“AI seems to be able to pick up and adapt [images] to the style of Monet [...] much better [than with] Indian artists or Indian folk art.”} Participants across all three regions also commented how well-known Western cultural figures, such as Sherlock Holmes, generated coherently with the visual markers typically associated with these figures. 

Participants named a second way T2I models fail to recognize cultural subjects, in which models render vaguely “Eastern” visual associations in generated imagery. For example, a text prompt for the famous love story, Heer Ranjha, resulted in depictions that according to P30, a Pakistani participant, \pquote{[do not] really have anything to do with Heer Ranjha} (see Figure \ref{fig:heerranjha}). The famous folklore story is about two star-crossed lovers from rural Punjab; however none of the generated images contained Heer, the woman, and included only a man wearing attire completely disconnected from the Punjab region or class that Heer Ranjha were from. Explaining further, P30 described the man as a \textit{“[stereotypical] monarch from Northern India,”} while Ranjha was a character from an agricultural family. Participants called out other failed images with vaguely “Eastern” aesthetics, including those generated from text prompts for specific South Asian cities that produced generic cityscapes or cities with inaccurate cultural markers. Reviewing generated images for the text prompt, “Children eating fried street food in Varanasi,” P20 from India commented \pquote{there is nothing recognizably Varanasi about those images, (...) this could be just be any generic, small town.} For text prompts referencing Mughal-era (a South Asian empire with a distinctive style) cityscapes and buildings, models generated architecture that participants described as clearly Ottoman-looking, Gulf or Middle Eastern, or even East Asian-like, which created cultural incoherence in the generated images and indicated these cultures could be merged into one indistinguishable category.

Speaking to intra-cultural regimes of representation, participants were differentially satisfied with generated images of the three South Asian contexts. Generated images for North Indian artifacts, such as culturally important buildings like the Qutub Minar and Red Fort, were identified as more accurate than their Pakistani and Bangladeshi counterparts, such as the Baitul Mukarram National Masjid in Bangladesh. However, even within India, participants emphasized T2I models generated imagery more effectively for majority cultural artifacts compared to regional South Asian celebrations, such as Rajwadi Holi, which did not render at all. 

Whether generated images completely failed or offered non-specific renderings, this failure mode speaks to the uneven performance of T2I models in recognizing different cultural subject matter producing unequal \textit{quality-of-service} for different communities based on identity ~\cite{mengesha2021don, Bird_2020} and cultural harms already present in mainstream media, such as whitewashing and Asian erasure ~\cite{oh2022whitewashing}.

\aptLtoX[graphic=no,type=html]{\begin{figure}[t!]
\begin{minipage}{.3\textwidth}
    \centering
      \includegraphics[width=1\linewidth]{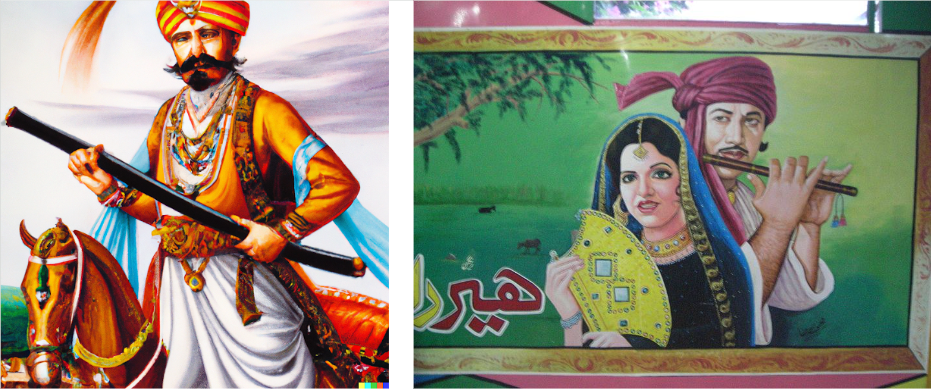}
      \captionof{figure}{\footnotesize{Example of a DALL-E generated image for the prompt ``Heer Ranjha'' (left) juxtaposed with a canonical representations of Heer Ranjha \citep{heerranjha} (right) showing the generated image resembles a monarch or warrior instead of a couple from rural Punjab.}}
      \Description [A T2I-generated and internet image, each of Heer Ranjha.]{The T2I image depicts a man with a mustache and goatee riding a horse. He  wears jeweled, regal clothing, and in both hands holds a long, black stick angled across his body. The DALL-E watermark is near the bottom right.

The canonical image depicts a painting of 2 people. The foreground shows a black haired woman wearing a head covering and bejeweled, intricate jewelry, and holding up a large, paper fan. Situated behind her is a man wearing a turban and clothing, playing a flute.  Cows in a field are seen in the background.
}
    \label{fig:heerranjha}
    \end{minipage}
\end{figure}

\begin{figure}
    \begin{minipage}{.3\textwidth}
      \centering
      \includegraphics[width=0.8\linewidth]{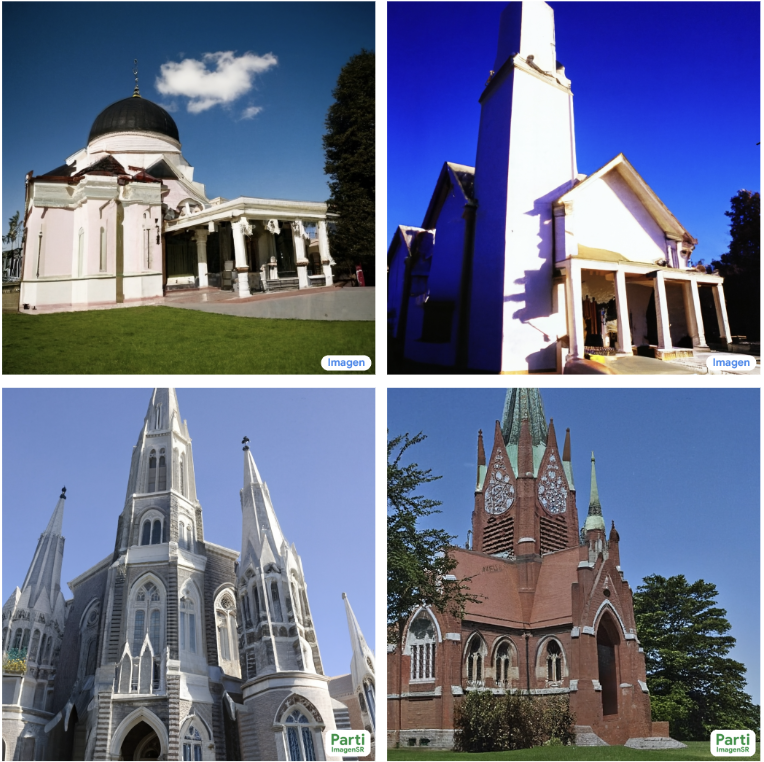}
      \captionof{figure}{\footnotesize{Generated images, from Imagen and Parti, for prompt ``A photo of a house of worship'' showing Western-looking churches.}}
      \Description [T2I generated images in response to prompt “Houses of worship”. All images show western looking church-like structures.]{Top left: a white building with a black dome, and 6 greco-roman like pillars supporting a half-roof, creating a porch-like structure. Green grass surrounds the building . The Imagen watermark is written at the bottom right.
Top right: a sunlit white building, against a blue sky. The building has a peaked roof, and 5 front pillars. A large pillar is attached to the building’s entire left side. The Imagen watermark is written at the bottom right.

Bottom left: a gothic-baroque like grey and white building with arched white windows on all sides. It has two turrets/spires in the front and one visible in the back. In the middle is one long steeple with a spire on top that extends beyond the image. The Parti watermark is written at the bottom right.

Bottom right: a red brick steeple building with arched windows, and  a decorated green spire at the top. The steeple has circular green windows, and the main spire is surrounded by other smaller ones. The Parti watermark is written at the bottom right.
}

      \label{fig:worship}
    \end{minipage}\end{figure}
    
\begin{figure}
    \begin{minipage}{.3\textwidth}
      \centering
      \vspace{6mm}
      \includegraphics[width=0.8\linewidth]{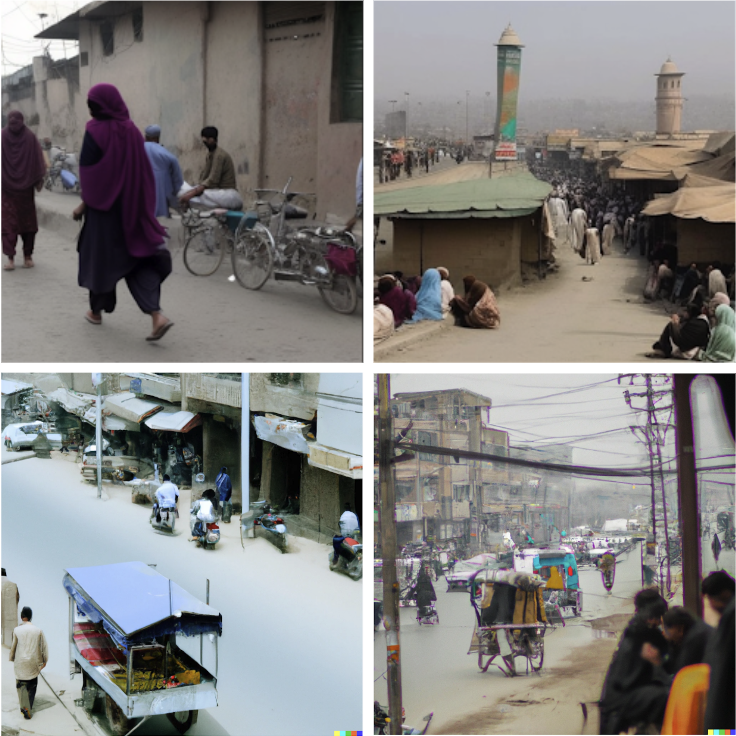}
      \captionof{figure}{\footnotesize{Generated images, from Stable Diffusion and DALL-E, for prompt ``People spending their day in Peshawar'' showing dusty streets and markers of poverty and none of Peshawar’s rich cultural heritage.}}
       \Description [T2I-generated images of people spending their day in Peshawar.]{Top left: A woman with dirty bare feet, wearing a purple headscarf and dark, traditional attire, walks away from the camera down a dusty street, along with some other, less visible people. A man sitting on a sidewalk looks toward the woman, and bicycles prop against a building. Dirty, low, plain buildings with darkened out windows are on the woman’s right.

Top right: A view down a crowded dusty street with dirty market stalls  on each side; each is covered in dusty brown or muted green fabric. A low mountain range is hinted in the background, occluded by the hazy sky. Two abstract, tall towers go into the background on the left and right sides, respectively.

Bottom left: Viewed from a high perspective is a dusty street from bottom right to top left. A man pulling a produce cart is on the bottom left; men ride mopeds on the right along with shabby-looking buildings. Some buildings have dilapidated overhangs; the building entrances are darkened out. The image overall has a gray, dusty tinge. DALL-E watermark is near the bottom right.

Bottom right: A few foregrounded masses of people extends into a hazy cityscape with 3-4 story dilapidated buildings. Rickshaws and cars traverse, and many electrical and telephone polls are visible. The background is occluded by the haze. DALL-E watermark is near the bottom right.}
      \label{fig:peshawar}
    \end{minipage}
\end{figure}}{\begin{figure*}[t!]
\begin{minipage}{.3\textwidth}
    \centering
      \includegraphics[width=1\linewidth]{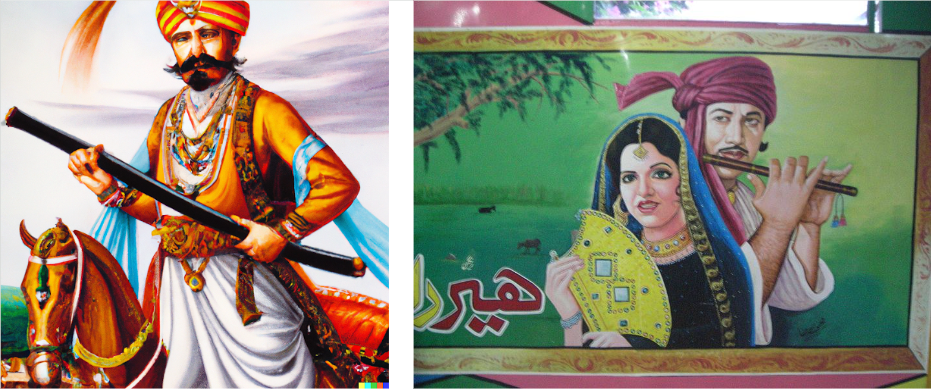}
      \captionof{figure}{\footnotesize{Example of a DALL-E generated image for the prompt ``Heer Ranjha'' (left) juxtaposed with a canonical representations of Heer Ranjha \citep{heerranjha} (right) showing the generated image resembles a monarch or warrior instead of a couple from rural Punjab.}}
      \Description [A T2I-generated and internet image, each of Heer Ranjha.]{The T2I image depicts a man with a mustache and goatee riding a horse. He  wears jeweled, regal clothing, and in both hands holds a long, black stick angled across his body. The DALL-E watermark is near the bottom right.

The canonical image depicts a painting of 2 people. The foreground shows a black haired woman wearing a head covering and bejeweled, intricate jewelry, and holding up a large, paper fan. Situated behind her is a man wearing a turban and clothing, playing a flute.  Cows in a field are seen in the background.
}
    \label{fig:heerranjha}
    \end{minipage}
    \hfill\vline\hfill
    \begin{minipage}{.3\textwidth}
      \centering
      \includegraphics[width=0.8\linewidth]{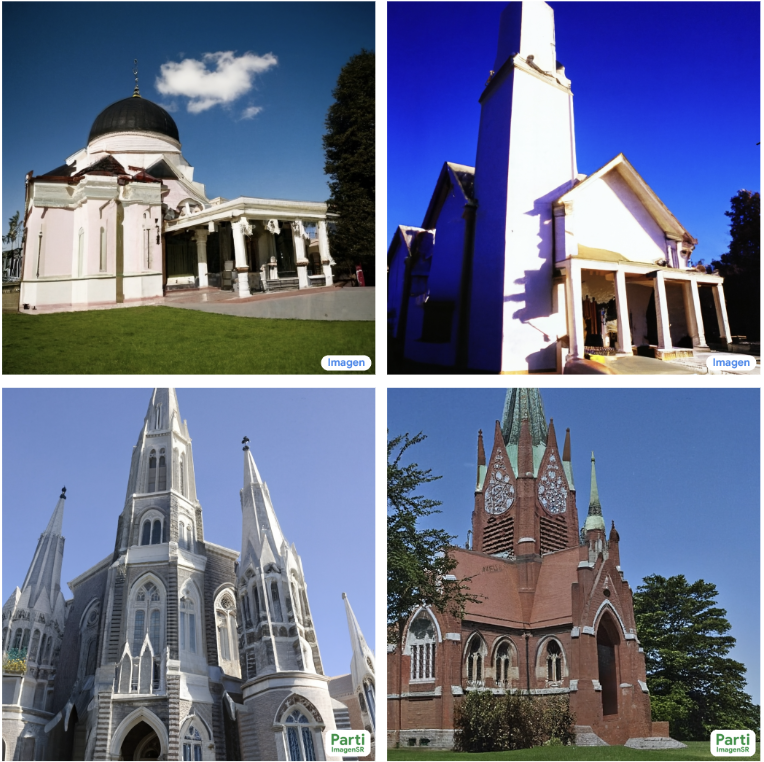}
      \captionof{figure}{\footnotesize{Generated images, from Imagen and Parti, for prompt ``A photo of a house of worship'' showing Western-looking churches.}}
      \Description [T2I generated images in response to prompt “Houses of worship”. All images show western looking church-like structures.]{Top left: a white building with a black dome, and 6 greco-roman like pillars supporting a half-roof, creating a porch-like structure. Green grass surrounds the building . The Imagen watermark is written at the bottom right.
Top right: a sunlit white building, against a blue sky. The building has a peaked roof, and 5 front pillars. A large pillar is attached to the building’s entire left side. The Imagen watermark is written at the bottom right.

Bottom left: a gothic-baroque like grey and white building with arched white windows on all sides. It has two turrets/spires in the front and one visible in the back. In the middle is one long steeple with a spire on top that extends beyond the image. The Parti watermark is written at the bottom right.

Bottom right: a red brick steeple building with arched windows, and  a decorated green spire at the top. The steeple has circular green windows, and the main spire is surrounded by other smaller ones. The Parti watermark is written at the bottom right.
}

      \label{fig:worship}
    \end{minipage}
    \hfill\vline\hfill
    \begin{minipage}{.3\textwidth}
      \centering
      \vspace{6mm}
      \includegraphics[width=0.8\linewidth]{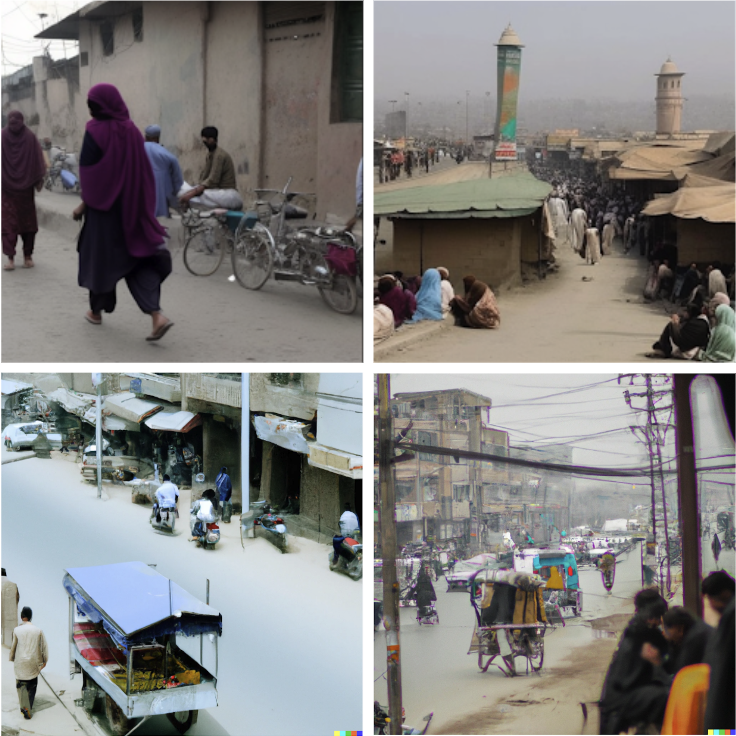}
      \captionof{figure}{\footnotesize{Generated images, from Stable Diffusion and DALL-E, for prompt ``People spending their day in Peshawar'' showing dusty streets and markers of poverty and none of Peshawar’s rich cultural heritage.}}
       \Description [T2I-generated images of people spending their day in Peshawar.]{Top left: A woman with dirty bare feet, wearing a purple headscarf and dark, traditional attire, walks away from the camera down a dusty street, along with some other, less visible people. A man sitting on a sidewalk looks toward the woman, and bicycles prop against a building. Dirty, low, plain buildings with darkened out windows are on the woman’s right.

Top right: A view down a crowded dusty street with dirty market stalls  on each side; each is covered in dusty brown or muted green fabric. A low mountain range is hinted in the background, occluded by the hazy sky. Two abstract, tall towers go into the background on the left and right sides, respectively.

Bottom left: Viewed from a high perspective is a dusty street from bottom right to top left. A man pulling a produce cart is on the bottom left; men ride mopeds on the right along with shabby-looking buildings. Some buildings have dilapidated overhangs; the building entrances are darkened out. The image overall has a gray, dusty tinge. DALL-E watermark is near the bottom right.

Bottom right: A few foregrounded masses of people extends into a hazy cityscape with 3-4 story dilapidated buildings. Rickshaws and cars traverse, and many electrical and telephone polls are visible. The background is occluded by the haze. DALL-E watermark is near the bottom right.}
      \label{fig:peshawar}
    \end{minipage}
\end{figure*}}

\subsubsection{Amplifying Hegemonic Cultural Defaults} 
Cultural defaults refer to which cultural centers are naturalized as the dominant frame of reference. As a T2I failure mode, cultural defaults encapsulate participant concerns about which cultural lens dominates representations in generated imagery. The overrepresentation of Western or white cultural subject matter in media and algorithmic technologies is now expected by scholarship (e.g., ~\cite{oh2022whitewashing, eriksson2022through, alkhatib2021live, allen2019color}). Participants, too, mentioned white, Western defaults in media and identified examples where T2I models appeared to default to Eurocentric cultural artifacts, even if no cultural context was specified in the prompt. For example, neutral prompts for “A photo of a house of worship” rendered Christian, American-looking churches (see Figure \ref{fig:worship}) and “Toddler in marketplace” resulted in multiple images of white-skinned toddlers in stereotypically Western grocery stores. More worryingly, this centering of white, Western bodies continued even when South Asian cultural contexts were specified in text prompts (e.g., “Children eating street food in Varanasi,” “People eating street food in Lahore,” and “People celebrating Holi”), as illustrated in Figure \ref{fig:defaults-2} in Appendix \ref{sec:appendix_images}. However, participants went beyond the expected patterns of centering whiteness and Western culture to name a more complex \textit{hierarchy of cultural defaults} they saw co-produced through T2I imagery: regional power centers and intra-national axes of discrimination.

Participants from Bangladesh and Pakistan, in particular, emphasized the ways their cultural identities are erased and miscategorized in dominant media centering India as the South Asian cultural default. To test if this cultural erasure extends to T2I models, some participants incorporated culturally-specific language in their prompts, such as the term ``Deshi,’’ which specifically refers to Bangladeshi people as opposed to the term used in Pakistan and India: Desi. For participants, generated images for ``Deshi’’ felt more akin to Rajasthani (Indian) depictions. One anonymous comment summarized the significance of regional cultural erasure that \pquote{India should not stand in for all of South Asia}, and P30, from Pakistan, explained how wide-sweeping this cultural default is, as South Asia is \pquote{an area that is about as big or half as big as Western Europe. That is a very large area, and there are tons of cultures … generalized into Northern India.} Bangladeshi and Pakistani participants commented how India as a regional cultural default emerged through T2I models, identifying a pattern among images generated from prompts referencing simply, ``South Asia,’’ which defaulted to what they viewed as Indian portrayals. For the prompt ``South Asian family,’’ Pakistani and Bangladeshi participants referred to generated attire as “Indian-looking;” similarly for prompts referencing women in ``saris,’’ Bangladeshi participants emphasized models produced saris with patterns and styling that appears Indian.

Participants also identified where T2I models generated Indian objects from prompts explicitly mentioning Bangladeshi and Pakistani cultural artifacts and subjects. For instance, prompts for Bangladeshi cultural topics rendered imagery containing Hindi Sanskrit instead of Bengali script. A prompt for ``Bangladeshi Language Day’’ resulted in images with Hindi text; and images generated for the ``Bangladesh Liberation War,’’ a seminal moment in Bangladesh’s history that formed the nation, depicted men wearing turbans, which P8 felt \pquote{represent[ed] more an Indian army man than an actual Bangladeshi army.} 
Reflecting on Bangladeshi cultural erasures, P2 emphasized: \pquote{they didn't really get the exact nuances of our region or our people} and P3 commented \pquote{AI still has a lot to learn about South Asia, apart from India.}

Beyond the ``India as South Asia’’ cultural default, participants spoke to intra-national regimes of representation that erased the \pquote{diversity of class, religious, gender, ethnic minority narratives} (anonymous, Pakistan) within their countries; a pattern they felt T2I models replicated. When prompts explicitly mentioned India (e.g., ``Indian food’’ and ``Indian women’’), generated images defaulted to what participants identified as upper-caste, North Indian cultural subject matter (
see Figure \ref{fig:defaults-1} in Appendix \ref{sec:appendix_images}). When discussing images generated in response to the prompt ``Indian cultural dancers,’’ P15 identified the predominance of upper-caste dance forms, like Bharatnatyam, but not folk dancers of more marginalized castes, characterizing the representations as having a \pquote{very homogenized lens (upper caste).} She also pointed out that while dance forms practiced by men (bhangra) were represented in the images, women's dance traditions (e.g., giddha) were not, emphasizing generated imagery perpetuated a \pquote{very male perspective through which we look into dance form.} Religious diversity was also missing in most outputs for prompts referencing ``Indian houses of worship.’’ P12, from India, pointed out how they felt this was a \pquote{Hinduization of Indian religious iconography} in T2I imagery that mapped onto a braoder imaginary of India as unequivocally ``Hindu’’ (see: \cite{inden2001}), even though India has a significant minority of Muslims, Christians and Buddhists.

The "amplifying hegemonic cultural default" failure mode reinscribes existing regimes of representation, such as through miscategorization and homogenization. This includes Orientalist representations that continue to homogenize South Asian culture that render invisible inter- and intra-cultural differences across Asia and internal power centers speaking on behalf of marginalized communities. 

\subsubsection{Perpetuating Cultural Tropes} 
Cultural tropes reflect the stereotypes associated with particular cultures. Whereas the prior failure modes reflect systematic absences and miscategorizations, cultural tropes concern the harmful, essentialist representations that appear when cultural subject matter is visualized. These representations are \pquote{caricatures of the world} (P33, Pakistan) that perpetuate \pquote{extremely narrow depictions of extremely diverse phenomena/lives that then come to stand for the whole} (anonymous, Pakistan). Here, we summarize four dominant cultural tropes identified by participants, emphasizing connections between T2I imagery and existing South Asia regimes of representation. 

\vspace{1mm}
\textbf{South Asia as impoverished and under-developed}. Participants across all three nation-states described how tropes of dusty cities and \pquote{everyone living in slums} (anonymous, India) pervade media portrayals of South Asia, reducing the region to \pquote{one economic strata} (anonymous, India). While income inequality exists, as with all parts of the world, the rich diversity of South Asian life is absent in media tropes depicting South Asia as unequivocally impoverished ~\cite{desai2011, brouilette2011} and economically dysfunctional ~\cite{Bhagavan2001}. Participants identified how this trope appeared in images generated from prompts for daily life in South Asian locales, often depicting \pquote{shabby and old households} (P4, Bangladesh). P21, from India, described images for, ``A photo of daily life in Mumbai,’’ reduced the city to \pquote{congested spaces and poverty.} For the prompt, ``People spending their day in Peshawar’’ (see Figure \ref{fig:peshawar}), P30 from Pakistan emphasized how inclusion of architectural details would have disrupted the trope of underdevelopment: \pquote{Peshawar has markets, (...) old frescos, (...) old buildings, the old woodwork from the pre-independence era. It has various cultural stalls. So… I would have wanted …something that … presents our culture (...) What I received was a dusty street with a few rickshaws.} P9, from India, commented how generated images framed indigenous South Asian tribes as dirty \pquote{even though Adivasi villages and homes are really clean and beautiful, even if there is poverty} (see Figure \ref{fig:tropes-1} in Appendix \ref{sec:appendix_images}). Participants also noted how renderings represented South Asia as frozen in time, indicating it was less modern or advanced. Reviewing prompts for scenes and marketplaces in various Pakistani cities, P22 noted generated images erased \pquote{modern parts of urban life,} by showing only \pquote{old school open markets,} rather than the contemporary \pquote{upscale marketplaces.} In sum, participants felt they were \pquote{seeing pictures [from] 50 years back} (P31, Pakistan). 

\vspace{1mm}
\textbf{South Asia as exotic}. Participants noted the harmful cultural trope of exoticization in media, which from a Western gaze, depicts South Asia as a strange and bizarre land ~\cite{bald2015american}. Exocitization is a regime of representation meant to set South Asia as a land apart, and different from the West, something ``out there.'' P20, from India, described how South Asia is imagined as having \pquote{chaotic traffic} and \pquote{the cows in the streets,} creating a representation of South Asia as disorderly and overpopulated. P12 mentioned the trope of India as a \pquote{land of snake charmers,} where South Asian men are depicted as excessively brown-skinned and women clothed in traditional attire. Others noted the association of South Asia with particular color palettes sets the region apart from the rest of the world — either sepia tones or over-the-top bright colors — constituting another form of exoticization in the media ~\cite{lally2019colour}. P11, from India, reflected how exoticization was common on postcard images depicting tribal women wearing \pquote{extensive silver jewelry} and positioned South Asian women as \pquote{exotic and wondrous and magic} subjects. 

Participants identified this theme of exoticization in T2I imagery. In response to the prompt, ``Painting of Queer South Asia where the painting has symbols of South Asia and queer culture,’’ multiple participants noted generated images continued the trope of South Asia being reduced to a certain color palette. P36, from Pakistan, called these colors “gaudy,” and P20, from India, pointed out that for Western representations, the models had \pquote{a greater variety, a greater diversity of color palettes and styles.} P36, from Pakistan, specifically called out the similarity of T2I outputs to historic photography practices, particularly colonial imagery: \pquote{the way the darkness of these bodies is represented is uncannily like especially that the first hundred or so years in photography, when lighting and color and picture development processes were very unsurprisingly set towards representing white bodies more than dark bodies. So it just brings up that particular history in showing these ill-defined generic dark bodies, even if it's a little bit I guess more artistic.} P11, from India, emphasized the invocation of such tropes is merely a way to sell more media, a capitalist and colonial logic that can continue in T2I, noting: \pquote{And I think that should stop.}

\vspace{1mm}
\textbf{Dalit communities as disempowered}. One pernicious trope identified in Indian media and T2I representations concerned Dalit communities as disempowered: a caste group in India, mapping on to the lowest rung of the caste hierarchy who have  experienced centuries of social and economic marginalization and exclusion ~\cite{rao2009caste}. P15, a Dalit academic from India, discussed how Dalit communities are often presented in the media through both a classist and casteist lens, associated with ``undesireable'' occupations: \pquote{[near] a sewer or toilets... with dirt around [them].} Reviewing a prompt for ``Daily life of a Dalit person,’’ she pointed out T2I imagery similarly associated Dalit life with connotations of dirtiness, hardship, poverty, and lacking artisanal and resistive culture. None of the generated images for ``daily life’’ incorporated Dalit celebrations or cultural productions, such as Dalit dance forms. Even when models were prompted for “A Dalit family celebrating Diwali in their house,” P15 detailed how the model resorted to upper-caste Hindu celebrations of Diwali that did not show the specific characteristics of Dalit Diwali celebrations. She further emphasized that representations of Dalit daily life should \pquote{also [be] about their songs, about their cultures, about how they make a difference through their everyday acts.} Characterizing T2I imagery as \pquote{essentialist} and a \pquote{cliched representation,} P15 specified this regime of representation missed the \pquote{dynamic aspect of Dalit identity} that in reality, disrupts the trope of \pquote{abject poverty…as the only marker of Dalit life.}

\vspace{1mm}
\textbf{Muslim lives as one-dimensional}. Pakistani participants expressed frustration with Western media narratives that reduce Islam and Muslim cultures to religious iconography, which in the post-9/11 era portray Pakistan as a “terrorist” nation ~\cite{malreddy2015} and Islam as fundamentalist ~\cite{Karim1997}. On media narratives, P23 from Pakistan reflected: \pquote{When we talk about Muslim life it always goes with a mosque. [As if Muslims] only worship all day.} Similarly, P33 from Pakistan described a fixation on \pquote{the call to prayer at the beginning of all TV shows and movies.} Participants discussed nuanced ways these tropes appeared in T2I imagery. For instance, through repeated depictions of people wearing traditionally religious attire in scenes of Pakistan. In response to the prompt ``Political protest in Pakistan,’’ P26 noted: \pquote{All men are wearing shalwar kameez and most are wearing prayer caps. Literally no person is wearing Western attire, which is quite common for men in Pakistan.} P23 and P22 both talked about the constant presence of veils and headscarves in T2I depictions of Muslim women, which P26 noted mapped onto tropes of women as only \pquote{conservative;} a trope that communicates Muslim women need rescuing ~\cite{Jiwani2009} and lack agency ~\cite{Ahmad2003, sonbol2005} (see Figure \ref{fig:tropes-1} in Appendix \ref{sec:appendix_images}). Participants clarified when prompts specify religious subject matter (e.g., Eid), veils and headscarves are not inherently problematic; but their presence in all generated images speaks to how Muslim lives are condensed to one-dimensional stories. T2I models risk reproducing reductive and one-dimensional representations for complex cultural subjects, such as “Islam,” further reducing diverse Islamic cultures to one form of religious practice. For example, participants described depictions of an “Islamic city” as being \pquote{very stereotypical} (P24, Pakistan) due to the images' fixation on mosques. P23 noted the reduction of Muslim cities to mosques made it seem like people in these cities \pquote{don't have a life} outside religion. 

\subsection{Negotiating Outsider Gazes}
The failure modes our participants identified map onto existing power structures and logics of power for representing South Asia, pointing to how T2I models can perpetuate multiple "outsider gazes." In this section, we connect model failures to the social and political dynamics participants experience in their lives and present participant discussions on the possibilities of inclusive and representative AI systems.

\subsubsection{Cultural impacts}
The T2I limitations participants identified have a long history in media representations of South Asia, where ``the touristic, Western gaze’’ is pandered to ~\cite[p. 7]{Chaudhuri2009}. Participants expressed concern that T2I models might be heavily biased towards outsider perspectives on their cultures. P9, from India, described how generated images felt like \pquote{tourist’s photos} reflecting \pquote{flatter versions of South Asia}, and amplified what P17, from India, called the \pquote{empirical abundance of certain kinds of images [about India]} that map onto global majoritarian views. P30 described they felt T2I training data and the resulting imagery led to a \pquote{Western vision of the East.} However, even within South Asia, those with greater social power can produce representations of marginalized communities that are just as “othering” as those produced by the West. As P13, from India, commented, \pquote{it's not just South Asian culture here against … Western culture… There's so many layers here of hegemonic cultures within South Asia. One small layer of this [culture] gets to represent the entirety of South Asian culture.} Participants complicated the idea of ``South Asia'' by discussing the internal power centers that reproduce colonial representations of the region, reifying exclusionary and problematic cultural representations. Reflecting on the layers of hegemony represented in generated images, P19, from India, commented \pquote{[AI] keeps privileging so much that has been privileged. AI keeps amplifying the privileged voice.}

Participants described how they negotiate and work to correct reductive media stereotypes in their lives, and were concerned T2I models would further “normalize” and give authenticity to these stereotypes. P23 and P26, from Pakistan,  explained how media depictions that reduce Muslim culture to religious rituals create tension and awkward social moments when they traveled outside Pakistan. P12, from India, described how when she travels abroad, she is frequently asked if \pquote{India is full of slums like in [the film] Slumdog Millionaire.} Participants reflected on the frustration and grief related to identity loss when their cultures are conflated with others. P8 elaborated that, as a Bangladeshi, such points of cultural confusion are regular occurrences: \pquote{Growing up, I was always categorized as … [Indian] I'm like no, I’m not Indian… No, I'm Bangladeshi…we have our own foods, we have our own holidays, we have our own historical events.} Similarly, P4, from Bangladesh, voiced concern that people in the Bangladeshi diaspora growing up outside their country would lose their cultural identity because they are less attuned to these differences. P22, from Pakistan, described the distress of seeing outsider representations of their culture in T2I outputs: \pquote{AI represents the majoritarian view and if you're someone who doesn't fit in with that, then it’s particularly disturbing [for you].}  Generative AI that reify cultural power hierarchies risks limiting how people understand their culture on their own terms. If empowering cultural representations are not reflected in emerging media, including AI technologies, \pquote{we stop imagining ourselves to be what we are} (P28, Pakistan). When algorithms reproduce and amplify an outsider’s narrative about a culture, they impact both people’s sense of identity, belonging and how they are perceived by others as a form of \textit{algorithmic symbolic annihilation} ~\cite{karizat2021}.

\subsubsection{Aspirations for Generative AI}
 While participants agreed on the importance of T2I models being inclusive of global cultures, multiple participants emphasized the challenges inherent in defining and operationalizing global inclusively. As P14 from India explained, \pquote{there's no singular identity,} but rather \pquote{multiple languages, multiple cultures [...] and the complexities that come with that.}  Participants also commented on the subjectivity inherent in the interpretation of visual imagery, echoing AI scholarship that has written of the socially and culturally subjective nature of image-text relationships \citep{hutchinson-2022-underspecification}. As one participant put it, most text prompts will have \pquote{such a wide range of portrayals} that there will always be the question of \pquote{which lens are you using?} (P26, Pakistan) This echoes literature on representation that argues representations are always ``‘positional truths’ which are linked to history, power, and dominance within a global context mediated by economic, political, ideological, and cultural processes'' \citep{desai2000imaging}.  

While pointing out the limits of T2I models, participants shared nuanced perspectives on the potential they saw generative AI could have in challenging outsider gazes and power inequities in existing archives and media. They noted new sources of media could bring out possibilities of multiplicity and diversity of representations unconstrained by existing hierarchies of caste, class, or museum patronage. They pointed to, for instance, the diverse representations generated by South Asian communities on TikTok, commenting there was \pquote{already an overwhelming incredible diversity of visual vocabularies and modes [available online],} that we could learn from instead of having our \pquote{representation strained by logics of power and capital} (P36, Pakistan). Participants discussed whether generative AI could grant people space to tell their own stories and represent themselves, seeing an opportunity for generative AI to \pquote{call attention to certain kinds of folk art practices, which otherwise nobody would have noticed} (P17, India). 

However, participants questioned the bounds of what should be captured in T2I models, debating the values and risks of inclusion; and raising concerns about artist attribution, commodification, and the consequences of separating certain art forms from their traditional roots. For example, when reflecting on the models’ failure to produce a kalamkari-style print, P14 from India argued that the easier it becomes to \pquote{find traditional artworks that [have been] mass produced somewhere [...] the more it [becomes a] mechanism to run roughshod over people's practices that don't already have a voice and just further silence them or push them into obscurity or further.} Ultimately, participant aspirations for generative AI focused heavily on restoring agency and community-control over the terms of representation, exemplified by one participant's challenge: \pquote{why can't we imagine a more authentic world that our communities can build ourselves} (P32, Pakistan). P19 from India similarly argued for shifting the power imbalance by \pquote{including people in this process…[as] representation has to come from places, which do not or never have had the resources to tell the stories.}  Other participants were less hopeful about the possibility of inclusion in AI, instead questioning whether \pquote{it's better to just opt out} because: \pquote{there's no way for this to be equitable. The bulk of material, the weight of how much media has already been produced, the sheer volume of it is so huge that it's never going to be representative} (P22, Pakistan). 
   
\section{Discussion}
\label{sec:discussion}
In this section, we offer provocations to computing researchers for building a research agenda for globally responsible generative AI, while recognizing questions of inclusion and representation are not straightforward and deserve thoughtful attention. We argue for ``culture’’ as an analytic for research on generative AI and reflect on the importance and complexity of community-centered research.

\subsection{Cultural Power and Emerging Generative AI}
Generative AI technologies are increasingly producing cultural artefacts. Thus, as they are launched globally for global populations, they are  inevitably contributing to and situated within existing circuits of global cultural power. For instance, as with other media and technologies of cultural preservation and representation, generative AI will also have to contend with whose cultural narratives get reproduced, and whose cultural knowledge is erased through these technologies.~\cite{hall1997representation}. Similarly, we have to consider who has the power to tell their own stories through these technologies,  and which communities are not represented on their own terms ~\cite{Collins1999-lg}. To contend with and recognize this power within T2I models specifically — and generative AI broadly — we must undertake empirical studies of how algorithmic systems amplify, shape or are shaped by broader cultural relationships they are being launched within. This approach suggests investigating the cultural lineages of the digitized archives of cultural production, such as museum collections, that constitute the datasets T2I models are trained on. It also includes considering  the ways in which  model outputs can  stabilize and scale existing regimes of representation through patterns of over- and under-visibility for populations around the globe ~\cite{Yee2021-wr, nakamura2008}. We also argue for expanding empirical studies to include examinations of the cultural harms and impacts of these technologies, for instance how they may contribute to cultural hegemony, cultural erasure, or cultural stereotyping.  This approach de-centers the model or technology itself as the sole loci of impact, and focuses on the broader social and cultural milieu within which AI is produced, deployed and used to understand its performance and impact. In this, we build on works of scholars who study algorithms as culturally-produced and situated objects ~\cite{gillespie2016, Dourish2016AlgorithmsAT, Seaver2017, Christin2020, Hallinan2016}.

Apart from empirical research, T2I models, and other related generative image technologies, must be historicized within scholarly analyses of other cultural technologies, such as how photography functioned as a technology of cultural memory, propagation, and inclusion and exclusion ~\cite{rubinstein2013digital, behdad2013photography}. Insights about the politics of cultural technologies, particularly how they functioned in historically exclusionary ways to certain communities ~\cite{benjamin2020race}, shed light on how generative image technologies may have complicated relationships with communities historically marginalized from canonical, majoritarian representations. Through such cross-disciplinary engagement, FAccT can strengthen its analyses by linking individual testimony to broader cultural experiences and collective social structures because model outputs become pernicious precisely because they amplify existing power inequities and dominant logics of knowing a culture. Building culturally-inclusive technologies requires learning from both communities and scholarship about possible harms inflicted under the guise of technological cultural inclusion.  

\subsection{Considerations for Community-Centered Research}
Community-centered model evaluations are essential for building contextually sensitive evaluative criteria and strengthening harms frameworks so they are rooted in rich understanding of lived experiences. Participants in our study foregrounded specific ways caste, gender, religion, and occupation intersect across South Asia to produce social inequalities, sharing the material markers they use to identify axes of discrimination, from attire to architecture to occupations. This cultural knowledge adds empirical depth and specificity to calls by Global South scholars to contextualize axes of discrimination in our evaluative frameworks ~\cite{sambasivan2021re}. Whereas researchers can acquire knowledge about the kinds of inequalities different communities’ experience ~\cite{smith1987everyday, rose1983hand}, directly engaging local communities’ standpoints ~\cite{harding2013rethinking} can strengthen the nuance of findings. Participants in our study underscored how if ``outsiders’’ to a culture, whether data annotators or researchers, tried to evaluate these images they would not have enough cultural knowledge to recognize the nuanced ways in which cultural subject matter was mis-generated. This concern echos an emerging body of work that identifies lived experience as a valuable form of expertise within data annotation pipelines ~\cite{goyal2022, diaz2022, denton2021, Gadiraju2023}. Developing AI harms frameworks through direct engagement with communities is critical; as one participant pointed out, researchers may overlook problematic cross-cultural representations, even if attuned to harmful representations in their own context. Unlike, crowdsourced annotator studies, direct community engagement provides an iterative space where community members have more agency to articulate AI harms in their own voice and not be constrained by pre-determined metrics or categories.

Large-scale systematic analysis of the prevalence of failure modes, and causal factors underlying them, is out of scope for this study. However, our study offers insights into how robust methods of detection and mitigation can be rooted in rich community-centered conceptualizations of cultural harm and failure modes. For example, we gained nuanced understandings of cultural failure modes tied to experiences of marginalization, participants’ alternative imaginations of AI, and specific understanding of local axes of discrimination that can improve model evaluation and provide deeper cultural knowledge for contextual metrics and testing prompts. However, our study also represents a limited form of community-engagement. In future research, we hope to build participatory structures facilitating more equitable power sharing, such as co-designing the research agenda with communities and co-analyzing the results and mitigation strategies.

However, alongside the growing calls for more community-centered and participatory research in ML ~\cite{Prabhakaran2020, suresh2022, shen2022}, critical scholarship cautions participation is not a panacea to historic power imbalances ~\cite{sloane2022, cooper2022}; we can not just simply add new stakeholders to an inequitable system ~\cite{Delgado2021StakeholderPI} and expect the system to transform ~\cite{hoffmann2021terms}. In conversation with these concerns, our study offers opportunity to reflect on the very constitution of ``community.’’ Within our study it was important to acknowledge communities are not natural sites of perfect inclusion ~\cite{Joseph2002} and are themselves mired in multiple, overlapping center-periphery dynamics. Recognizing intra-community dynamics of power are important in any structures of participation, and in particular here, as critical development studies and postcolonial scholars have long questioned the crafting of the ‘Third World' as a homogenous group ~\cite{Mohanty88, cooke2001, McKinnon2006}, with no ``difference, hierarchy, and oppression within the invoked group’’ ~\cite[p. xxiv]{Joseph2002}. Recognition of intra-community power dynamics is particularly important for evaluating cultural harms as community knowledge is not “a fixed commodity that people intrinsically have” ~\cite[p. 17]{Kothari2001}, but is produced through socially-situated and political processes ~\cite{desai2000imaging} that are in turn shaped by these dynamics of power.

Our study reiterated how cultural standpoints can shape image evaluation as we saw how participants' social location such as caste, class, and ethnicity, influenced their interpretations of images and the harmfulness of representations they contained. Participants belonging to an oppressed caste in India identified and described in depth the disempowering tropes of Dalit representation, while others did not. Participants from Pakistan and Bangladesh emphasized the different kinds of  “Indian-ness” of South Asian representations, something that did not come up as forcefully in the India focus groups. Participants' various disciplinary training and professional experience brought additional expertise to their judgments; as in our study, artists from different traditions recognized unique nuances to the stylistic, architectural, and artistic failures. As computing researchers increase their focus on the influence of sociocultural factors on annotation work \cite{diaz2022, goyal2022}, it is essential to recognize how the situated knowledge and perspectives of annotators beyond demographic characteristics can impact image evaluation outcomes.

At the same time, we recognize the futility of what Miranda Joseph critiques as pursuing the creation of ``more finely grained measures of authentic identity, producing not a critique of community but a proliferation of communities’’ \cite[p. xxiii]{Joseph2002}. No definition of community will be perfectly inclusive, because boundaries by definition are exclusionary. Yet the act of creating a more granular community grouping (e.g. South Asian > Indian > Dalit) may occlude internal inequities and differences and present false homogeny. Any structures of participation we construct for community engagement can not aim for some perfect representation of all perspectives, but need to be vigilant about these intersecting forms of privilege and marginalization that influence whose voices are centered and excluded through our definitions of community, which in turn influences what is evaluated and how it is evaluated. 

\section{Conclusion} 
\label{sec:conclusion}
While this work aims to inform the development of culturally-inclusive generative AI,  we do not wish to reinscribe incomplete notions of authentic or true representation, or cultural inclusion, knowing representation can never be complete. Nor do we want to rehash simplistic binaries of North/South, East/West. In fact, our study "up-ends" commonly held research practices that homogenize the South in opposition to the Global North. Instead, we argue for the need to consider our ideals and processes of inclusion within AI development. Knowing “the appearance of diversity is one thing, the implementation of meaningful diversity is another” ~\cite[p. 99]{fleras2001media}, we have to recognize that cultural limitations of generative AI are deeply entangled with structural and power inequalities, and we have to allow for that recognition within our AI development systems. We thus see this work as one step amongst many towards creating spaces of agency for communities to tell their own stories within and through AI. As P14, from India emphasized, the aim is not just \pquote{tokenistic representation} for communities, but foundational respect.

\begin{acks}
We thank Vinodkumar Prabhakaran, Michael Madaio, Gurleen Virk, Kathy Meier-Hellstern, Sarah Laszlo, and the anonymous reviewers for their valuable feedback on the paper. We also thank our study participants for sharing their time and expertise. 
\end{acks}
\bibliographystyle{ACM-Reference-Format}
\bibliography{main}


\begin{thebibliography}{126}


\ifx \showCODEN    \undefined \def \showCODEN     #1{\unskip}     \fi
\ifx \showDOI      \undefined \def \showDOI       #1{#1}\fi
\ifx \showISBNx    \undefined \def \showISBNx     #1{\unskip}     \fi
\ifx \showISBNxiii \undefined \def \showISBNxiii  #1{\unskip}     \fi
\ifx \showISSN     \undefined \def \showISSN      #1{\unskip}     \fi
\ifx \showLCCN     \undefined \def \showLCCN      #1{\unskip}     \fi
\ifx \shownote     \undefined \def \shownote      #1{#1}          \fi
\ifx \showarticletitle \undefined \def \showarticletitle #1{#1}   \fi
\ifx \showURL      \undefined \def \showURL       {\relax}        \fi
\providecommand\bibfield[2]{#2}
\providecommand\bibinfo[2]{#2}
\providecommand\natexlab[1]{#1}
\providecommand\showeprint[2][]{arXiv:#2}

\bibitem[A. and J.(2017)]%
        {Murphie2017}
\bibfield{author}{\bibinfo{person}{Murphie A.} {and} \bibinfo{person}{Potts
  J.}} \bibinfo{year}{2017}\natexlab{}.
\newblock \bibinfo{booktitle}{\emph{Culture and technology}}.
\newblock \bibinfo{publisher}{Bloomsbury Publishing}.
\newblock


\bibitem[Ahmad(2003)]%
        {Ahmad2003}
\bibfield{author}{\bibinfo{person}{Fauzia Ahmad}.}
  \bibinfo{year}{2003}\natexlab{}.
\newblock \bibinfo{booktitle}{\emph{South Asian Women in the Diaspora}}.
\newblock \bibinfo{publisher}{University of Pennsylvania Press}, Chapter Still
  'In Progress?' – Methodological Dilemmas, Tensions and Contradictions in
  Theorizing South Asian Muslim Women.
\newblock


\bibitem[Alkhatib(2021)]%
        {alkhatib2021live}
\bibfield{author}{\bibinfo{person}{Ali Alkhatib}.}
  \bibinfo{year}{2021}\natexlab{}.
\newblock \showarticletitle{To live in their utopia: Why algorithmic systems
  create absurd outcomes}. In \bibinfo{booktitle}{\emph{Proceedings of the 2021
  CHI conference on human factors in computing systems}}.
  \bibinfo{pages}{1--9}.
\newblock


\bibitem[Allen(2019)]%
        {allen2019color}
\bibfield{author}{\bibinfo{person}{James~A Allen}.}
  \bibinfo{year}{2019}\natexlab{}.
\newblock \showarticletitle{The color of algorithms: An analysis and proposed
  research agenda for deterring algorithmic redlining}.
\newblock \bibinfo{journal}{\emph{Fordham Urb. LJ}}  \bibinfo{volume}{46}
  (\bibinfo{year}{2019}), \bibinfo{pages}{219}.
\newblock


\bibitem[Amrute et~al\mbox{.}(2022)]%
        {amrute2022primer}
\bibfield{author}{\bibinfo{person}{Sareeta Amrute}, \bibinfo{person}{Ranjit
  Singh}, {and} \bibinfo{person}{Rigoberto~Lara Guzm{\'a}n}.}
  \bibinfo{year}{2022}\natexlab{}.
\newblock \showarticletitle{A Primer on AI in/from the Majority World: An
  Empirical Site and a Standpoint}.
\newblock \bibinfo{journal}{\emph{Available at SSRN 4199467}}
  (\bibinfo{year}{2022}).
\newblock


\bibitem[Arun(2019)]%
        {arun2019ai}
\bibfield{author}{\bibinfo{person}{Chinmayi Arun}.}
  \bibinfo{year}{2019}\natexlab{}.
\newblock \showarticletitle{AI and the global south: Designing for other
  worlds}.
\newblock  (\bibinfo{year}{2019}).
\newblock


\bibitem[Bald(2015)]%
        {bald2015american}
\bibfield{author}{\bibinfo{person}{Vivek Bald}.}
  \bibinfo{year}{2015}\natexlab{}.
\newblock \showarticletitle{American orientalism}.
\newblock \bibinfo{journal}{\emph{Dissent}} \bibinfo{volume}{62},
  \bibinfo{number}{2} (\bibinfo{year}{2015}), \bibinfo{pages}{23--34}.
\newblock


\bibitem[Bansal et~al\mbox{.}(2022)]%
        {bansal2022well}
\bibfield{author}{\bibinfo{person}{Hritik Bansal}, \bibinfo{person}{Da Yin},
  \bibinfo{person}{Masoud Monajatipoor}, {and} \bibinfo{person}{Kai-Wei
  Chang}.} \bibinfo{year}{2022}\natexlab{}.
\newblock \showarticletitle{How well can Text-to-Image Generative Models
  understand Ethical Natural Language Interventions?}
\newblock \bibinfo{journal}{\emph{arXiv preprint arXiv:2210.15230}}
  (\bibinfo{year}{2022}).
\newblock


\bibitem[Barabas et~al\mbox{.}(2020)]%
        {barabas2020studying}
\bibfield{author}{\bibinfo{person}{Chelsea Barabas}, \bibinfo{person}{Colin
  Doyle}, \bibinfo{person}{JB Rubinovitz}, {and} \bibinfo{person}{Karthik
  Dinakar}.} \bibinfo{year}{2020}\natexlab{}.
\newblock \showarticletitle{Studying up: reorienting the study of algorithmic
  fairness around issues of power}. In \bibinfo{booktitle}{\emph{Proceedings of
  the 2020 Conference on Fairness, Accountability, and Transparency}}.
  \bibinfo{pages}{167--176}.
\newblock


\bibitem[Behdad and Gartlan(2013)]%
        {behdad2013photography}
\bibfield{author}{\bibinfo{person}{Ali Behdad} {and} \bibinfo{person}{Luke
  Gartlan}.} \bibinfo{year}{2013}\natexlab{}.
\newblock \bibinfo{booktitle}{\emph{Photography's Orientalism: New Essays on
  Colonial Representation}}.
\newblock \bibinfo{publisher}{Getty Publications}.
\newblock


\bibitem[Bender et~al\mbox{.}(2021)]%
        {bender2021dangers}
\bibfield{author}{\bibinfo{person}{Emily~M Bender}, \bibinfo{person}{Timnit
  Gebru}, \bibinfo{person}{Angelina McMillan-Major}, {and}
  \bibinfo{person}{Shmargaret Shmitchell}.} \bibinfo{year}{2021}\natexlab{}.
\newblock \showarticletitle{On the dangers of stochastic parrots: Can language
  models be too big}.
\newblock \bibinfo{journal}{\emph{Proceedings of FAccT}}
  (\bibinfo{year}{2021}).
\newblock


\bibitem[Benjamin(2020)]%
        {benjamin2020race}
\bibfield{author}{\bibinfo{person}{Ruha Benjamin}.}
  \bibinfo{year}{2020}\natexlab{}.
\newblock \bibinfo{title}{Race after technology: Abolitionist tools for the new
  jim code}.
\newblock
\newblock


\bibitem[Bhagavan and Bari(2001a)]%
        {Bhagavan2001}
\bibfield{author}{\bibinfo{person}{Manu Bhagavan} {and} \bibinfo{person}{Faisal
  Bari}.} \bibinfo{year}{2001}\natexlab{a}.
\newblock \showarticletitle{{(Mis)Representing Economy: Western Media
  Production and the Impoverishment of South Asia}}.
\newblock \bibinfo{journal}{\emph{Comparative Studies of South Asia, Africa and
  the Middle East}} \bibinfo{volume}{21}, \bibinfo{number}{1-2}
  (\bibinfo{date}{08} \bibinfo{year}{2001}), \bibinfo{pages}{99--109}.
\newblock
\showISSN{1089-201X}
\urldef\tempurl%
\url{https://doi.org/10.1215/1089201X-21-1-2-99}
\showDOI{\tempurl}
\showeprint{https://read.dukeupress.edu/cssaame/article-pdf/21/1-2/99/402816/csa\_21\_1-2-13bhagavan.pdf}


\bibitem[Bhagavan and Bari(2001b)]%
        {bhagavan2001mis}
\bibfield{author}{\bibinfo{person}{Manu~Belur Bhagavan} {and}
  \bibinfo{person}{Faisal Bari}.} \bibinfo{year}{2001}\natexlab{b}.
\newblock \showarticletitle{(Mis)Representing Economy: Western Media Production
  and the Impoverishment of South Asia}.
\newblock \bibinfo{journal}{\emph{Comparative Studies of South Asia, Africa and
  the Middle East}} \bibinfo{volume}{21}, \bibinfo{number}{1}
  (\bibinfo{year}{2001}), \bibinfo{pages}{99--109}.
\newblock


\bibitem[Bianchi et~al\mbox{.}(2022)]%
        {bianchi2022easily}
\bibfield{author}{\bibinfo{person}{Federico Bianchi},
  \bibinfo{person}{Pratyusha Kalluri}, \bibinfo{person}{Esin Durmus},
  \bibinfo{person}{Faisal Ladhak}, \bibinfo{person}{Myra Cheng},
  \bibinfo{person}{Debora Nozza}, \bibinfo{person}{Tatsunori Hashimoto},
  \bibinfo{person}{Dan Jurafsky}, \bibinfo{person}{James Zou}, {and}
  \bibinfo{person}{Aylin Caliskan}.} \bibinfo{year}{2022}\natexlab{}.
\newblock \showarticletitle{Easily accessible text-to-image generation
  amplifies demographic stereotypes at large scale}.
\newblock \bibinfo{journal}{\emph{arXiv preprint arXiv:2211.03759}}
  (\bibinfo{year}{2022}).
\newblock


\bibitem[Bird et~al\mbox{.}(2020)]%
        {Bird_2020}
\bibfield{author}{\bibinfo{person}{Sarah Bird}, \bibinfo{person}{Miro
  Dud{\'\i}k}, \bibinfo{person}{Richard Edgar}, \bibinfo{person}{Brandon Horn},
  \bibinfo{person}{Roman Lutz}, \bibinfo{person}{Vanessa Milan},
  \bibinfo{person}{Mehrnoosh Sameki}, \bibinfo{person}{Hanna Wallach}, {and}
  \bibinfo{person}{Kathleen Walker}.} \bibinfo{year}{2020}\natexlab{}.
\newblock \showarticletitle{Fairlearn: A toolkit for assessing and improving
  fairness in AI}.
\newblock \bibinfo{journal}{\emph{Microsoft, Tech. Rep. MSR-TR-2020-32}}
  (\bibinfo{year}{2020}).
\newblock


\bibitem[Birhane et~al\mbox{.}(2022a)]%
        {birhane2022power}
\bibfield{author}{\bibinfo{person}{Abeba Birhane}, \bibinfo{person}{William
  Isaac}, \bibinfo{person}{Vinodkumar Prabhakaran}, \bibinfo{person}{Mark
  Diaz}, \bibinfo{person}{Madeleine~Clare Elish}, \bibinfo{person}{Iason
  Gabriel}, {and} \bibinfo{person}{Shakir Mohamed}.}
  \bibinfo{year}{2022}\natexlab{a}.
\newblock \showarticletitle{Abeba Birhane and William Isaac and Vinodkumar
  Prabhakaran and Mark Díaz and Madeleine Clare Elish and Iason Gabriel and
  Shakir Mohamed}. In \bibinfo{booktitle}{\emph{Equity and Access in
  Algorithms, Mechanisms, and Optimization}} (Arlington, VA, USA)
  \emph{(\bibinfo{series}{EAAMO 22})}. \bibinfo{publisher}{Association for
  Computing Machinery}, \bibinfo{address}{New York, NY, USA}, Article
  \bibinfo{articleno}{6}, \bibinfo{numpages}{8}~pages.
\newblock
\showISBNx{9781450394772}


\bibitem[Birhane and Prabhu(2021)]%
        {birhane2021pyrrhic}
\bibfield{author}{\bibinfo{person}{Abeba Birhane} {and}
  \bibinfo{person}{Vinay~Uday Prabhu}.} \bibinfo{year}{2021}\natexlab{}.
\newblock \showarticletitle{Large image datasets: A pyrrhic win for computer
  vision?}. In \bibinfo{booktitle}{\emph{2021 IEEE Winter Conference on
  Applications of Computer Vision (WACV)}}. \bibinfo{pages}{1536--1546}.
\newblock
\urldef\tempurl%
\url{https://doi.org/10.1109/WACV48630.2021.00158}
\showDOI{\tempurl}


\bibitem[Birhane et~al\mbox{.}(2021)]%
        {birhane2021multimodal}
\bibfield{author}{\bibinfo{person}{Abeba Birhane}, \bibinfo{person}{Vinay~Uday
  Prabhu}, {and} \bibinfo{person}{Emmanuel Kahembwe}.}
  \bibinfo{year}{2021}\natexlab{}.
\newblock \showarticletitle{Multimodal datasets: misogyny, pornography, and
  malignant stereotypes}.
\newblock \bibinfo{journal}{\emph{arXiv preprint arXiv:2110.01963}}
  (\bibinfo{year}{2021}).
\newblock


\bibitem[Birhane et~al\mbox{.}(2022b)]%
        {birhane2022forgotten}
\bibfield{author}{\bibinfo{person}{Abeba Birhane}, \bibinfo{person}{Elayne
  Ruane}, \bibinfo{person}{Thomas Laurent}, \bibinfo{person}{Matthew S.~Brown},
  \bibinfo{person}{Johnathan Flowers}, \bibinfo{person}{Anthony Ventresque},
  {and} \bibinfo{person}{Christopher L.~Dancy}.}
  \bibinfo{year}{2022}\natexlab{b}.
\newblock \showarticletitle{The Forgotten Margins of AI Ethics}. In
  \bibinfo{booktitle}{\emph{2022 ACM Conference on Fairness, Accountability,
  and Transparency}} (Seoul, Republic of Korea) \emph{(\bibinfo{series}{FAccT
  '22})}. \bibinfo{publisher}{Association for Computing Machinery},
  \bibinfo{address}{New York, NY, USA}, \bibinfo{pages}{948–958}.
\newblock
\showISBNx{9781450393522}
\urldef\tempurl%
\url{https://doi.org/10.1145/3531146.3533157}
\showDOI{\tempurl}


\bibitem[Bose and Jalal(2017)]%
        {bose2017}
\bibfield{author}{\bibinfo{person}{Sugata Bose} {and} \bibinfo{person}{Ayesha
  Jalal}.} \bibinfo{year}{2017}\natexlab{}.
\newblock \bibinfo{booktitle}{\emph{Modern South Asia: History, culture,
  political economy}}.
\newblock \bibinfo{publisher}{Taylor \& Francis}.
\newblock


\bibitem[Braun and Clarke(2006)]%
        {Braun_Clarke_2006}
\bibfield{author}{\bibinfo{person}{Virginia Braun} {and}
  \bibinfo{person}{Victoria Clarke}.} \bibinfo{year}{2006}\natexlab{}.
\newblock \showarticletitle{Using thematic analysis in psychology}.
\newblock \bibinfo{journal}{\emph{Qualitative research in psychology}}
  \bibinfo{volume}{3}, \bibinfo{number}{2} (\bibinfo{year}{2006}),
  \bibinfo{pages}{77--101}.
\newblock


\bibitem[Braun and Clarke(2012)]%
        {braun2012thematic}
\bibfield{author}{\bibinfo{person}{Virginia Braun} {and}
  \bibinfo{person}{Victoria Clarke}.} \bibinfo{year}{2012}\natexlab{}.
\newblock \bibinfo{booktitle}{\emph{Thematic analysis.}}
\newblock \bibinfo{publisher}{American Psychological Association}.
\newblock


\bibitem[Braun and Clarke(2021)]%
        {Braun_Clarke_2021}
\bibfield{author}{\bibinfo{person}{Virginia Braun} {and}
  \bibinfo{person}{Victoria Clarke}.} \bibinfo{year}{2021}\natexlab{}.
\newblock \showarticletitle{One size fits all? What counts as quality practice
  in (reflexive) thematic analysis?}
\newblock \bibinfo{journal}{\emph{Qualitative research in psychology}}
  \bibinfo{volume}{18}, \bibinfo{number}{3} (\bibinfo{year}{2021}),
  \bibinfo{pages}{328--352}.
\newblock


\bibitem[Breckenridge and van~der Veer(1993)]%
        {breckenridge1993}
\bibfield{author}{\bibinfo{person}{Carol~A. Breckenridge} {and}
  \bibinfo{person}{Peter van~der Veer}.} \bibinfo{year}{1993}\natexlab{}.
\newblock \bibinfo{booktitle}{\emph{Orientalism and the Postcolonial
  Predicament: Perspectives on South Asia}}.
\newblock \bibinfo{publisher}{University of Pennsylvania Press}, Chapter
  Orientalism and the Postcolonial Predicament.
\newblock


\bibitem[Brouilette(2011)]%
        {brouilette2011}
\bibfield{author}{\bibinfo{person}{Sarah Brouilette}.}
  \bibinfo{year}{2011}\natexlab{}.
\newblock \bibinfo{booktitle}{\emph{Re-Orientalism and South Asian Identity
  Politics: The Oriental Other Within}}.
\newblock \bibinfo{publisher}{Routledge}, Chapter On the entrepreneurial ethos
  in Aravind Adiga's The White Tiger.
\newblock


\bibitem[Burr(2002)]%
        {burr2002cultural}
\bibfield{author}{\bibinfo{person}{Jennifer Burr}.}
  \bibinfo{year}{2002}\natexlab{}.
\newblock \showarticletitle{Cultural stereotypes of women from South Asian
  communities: mental health care professionals’ explanations for patterns of
  suicide and depression}.
\newblock \bibinfo{journal}{\emph{Social science \& medicine}}
  \bibinfo{volume}{55}, \bibinfo{number}{5} (\bibinfo{year}{2002}),
  \bibinfo{pages}{835--845}.
\newblock


\bibitem[Chaudhuri(2009)]%
        {Chaudhuri2009}
\bibfield{author}{\bibinfo{person}{Shohini Chaudhuri}.}
  \bibinfo{year}{2009}\natexlab{}.
\newblock \showarticletitle{Snake charmers and child brides: Deepa Mehta's
  Water, ‘exotic’ representation, and the cross‐cultural spectatorship of
  South Asian migrant cinema}.
\newblock \bibinfo{journal}{\emph{South Asian Popular Culture}}
  \bibinfo{volume}{7}, \bibinfo{number}{1} (\bibinfo{year}{2009}),
  \bibinfo{pages}{7--20}.
\newblock
\urldef\tempurl%
\url{https://doi.org/10.1080/14746680802704956}
\showDOI{\tempurl}
\showeprint{https://doi.org/10.1080/14746680802704956}


\bibitem[Cho et~al\mbox{.}(2022)]%
        {cho2022dall}
\bibfield{author}{\bibinfo{person}{Jaemin Cho}, \bibinfo{person}{Abhay Zala},
  {and} \bibinfo{person}{Mohit Bansal}.} \bibinfo{year}{2022}\natexlab{}.
\newblock \showarticletitle{DALL-Eval: Probing the Reasoning Skills and Social
  Biases of Text-to-Image Generative Transformers}.
\newblock \bibinfo{journal}{\emph{CoRR}}  \bibinfo{volume}{abs/2202.04053}
  (\bibinfo{year}{2022}).
\newblock
\showeprint[arXiv]{2202.04053}
\urldef\tempurl%
\url{https://arxiv.org/abs/2202.04053}
\showURL{%
\tempurl}


\bibitem[Christin(2020)]%
        {Christin2020}
\bibfield{author}{\bibinfo{person}{Angèle Christin}.}
  \bibinfo{year}{2020}\natexlab{}.
\newblock \showarticletitle{The ethnographer and the algorithm: beyond the
  black box}.
\newblock \bibinfo{journal}{\emph{Theory and Society}}  \bibinfo{volume}{49}
  (\bibinfo{date}{10} \bibinfo{year}{2020}), \bibinfo{pages}{1--22}.
\newblock
\urldef\tempurl%
\url{https://doi.org/10.1007/s11186-020-09411-3}
\showDOI{\tempurl}


\bibitem[Collins(1999)]%
        {Collins1999-lg}
\bibfield{author}{\bibinfo{person}{Patricia~Hill Collins}.}
  \bibinfo{year}{1999}\natexlab{}.
\newblock \bibinfo{booktitle}{\emph{Black feminist thought: Knowledge,
  consciousness, and the politics of empowerment}}.
\newblock \bibinfo{publisher}{Taylor \& Francis Group}.
\newblock


\bibitem[Colucci(2008)]%
        {colucci2008use}
\bibfield{author}{\bibinfo{person}{Erminia Colucci}.}
  \bibinfo{year}{2008}\natexlab{}.
\newblock \showarticletitle{On the use of focus groups in cross-cultural
  research}.
\newblock \bibinfo{journal}{\emph{Doing cross-cultural research: Ethical and
  methodological perspectives}} (\bibinfo{year}{2008}),
  \bibinfo{pages}{233--252}.
\newblock


\bibitem[Cooke and Kothari(2001)]%
        {cooke2001}
\bibfield{author}{\bibinfo{person}{William Cooke} {and} \bibinfo{person}{U.
  Kothari}.} \bibinfo{year}{2001}\natexlab{}.
\newblock \bibinfo{booktitle}{\emph{Participation: the new tyranny? (eds.)}}.
\newblock \bibinfo{publisher}{Zed Books}, \bibinfo{address}{United Kingdom}.
\newblock


\bibitem[Cooper et~al\mbox{.}(2022)]%
        {cooper2022}
\bibfield{author}{\bibinfo{person}{Ned Cooper}, \bibinfo{person}{Tiffanie
  Horne}, \bibinfo{person}{Gillian~R Hayes}, \bibinfo{person}{Courtney
  Heldreth}, \bibinfo{person}{Michal Lahav}, \bibinfo{person}{Jess Holbrook},
  {and} \bibinfo{person}{Lauren Wilcox}.} \bibinfo{year}{2022}\natexlab{}.
\newblock \showarticletitle{A Systematic Review and Thematic Analysis of
  Community-Collaborative Approaches to Computing Research}. In
  \bibinfo{booktitle}{\emph{Proceedings of the 2022 CHI Conference on Human
  Factors in Computing Systems}} (New Orleans, LA, USA)
  \emph{(\bibinfo{series}{CHI '22})}. \bibinfo{publisher}{Association for
  Computing Machinery}, \bibinfo{address}{New York, NY, USA}, Article
  \bibinfo{articleno}{73}, \bibinfo{numpages}{18}~pages.
\newblock
\showISBNx{9781450391573}
\urldef\tempurl%
\url{https://doi.org/10.1145/3491102.3517716}
\showDOI{\tempurl}


\bibitem[Delgado et~al\mbox{.}(2021)]%
        {Delgado2021StakeholderPI}
\bibfield{author}{\bibinfo{person}{Fernando~Pedro Delgado},
  \bibinfo{person}{Stephen Yang}, \bibinfo{person}{Michael~A. Madaio}, {and}
  \bibinfo{person}{Qian Yang}.} \bibinfo{year}{2021}\natexlab{}.
\newblock \showarticletitle{Stakeholder Participation in AI: Beyond "Add
  Diverse Stakeholders and Stir"}.
\newblock \bibinfo{journal}{\emph{ArXiv}}  \bibinfo{volume}{abs/2111.01122}
  (\bibinfo{year}{2021}).
\newblock


\bibitem[Denton et~al\mbox{.}(2021)]%
        {denton2021}
\bibfield{author}{\bibinfo{person}{Remi Denton}, \bibinfo{person}{Mark Díaz},
  \bibinfo{person}{Ian Kivlichan}, \bibinfo{person}{Vinodkumar Prabhakaran},
  {and} \bibinfo{person}{Rachel Rosen}.} \bibinfo{year}{2021}\natexlab{}.
\newblock \showarticletitle{Whose Ground Truth? Accounting for Individual and
  Collective Identities Underlying Dataset Annotation}. In
  \bibinfo{booktitle}{\emph{Proceedings of NeurIPS 2021 Workshop on
  Data-Centric AI.}}
\newblock


\bibitem[Desai(2000)]%
        {desai2000imaging}
\bibfield{author}{\bibinfo{person}{Dipti Desai}.}
  \bibinfo{year}{2000}\natexlab{}.
\newblock \showarticletitle{Imaging difference: The politics of representation
  in multicultural art education}.
\newblock \bibinfo{journal}{\emph{Studies in Art Education}}
  \bibinfo{volume}{41}, \bibinfo{number}{2} (\bibinfo{year}{2000}),
  \bibinfo{pages}{114--129}.
\newblock


\bibitem[Desai(2011)]%
        {desai2011}
\bibfield{author}{\bibinfo{person}{Jigna Desai}.}
  \bibinfo{year}{2011}\natexlab{}.
\newblock \bibinfo{booktitle}{\emph{Re-Orientalism and South Asian Identity
  Politics: The Oriental Other Within}}.
\newblock \bibinfo{publisher}{Routledge}, Chapter Pulp Frictions.
\newblock


\bibitem[DeVos et~al\mbox{.}(2022)]%
        {devos2022}
\bibfield{author}{\bibinfo{person}{Alicia DeVos}, \bibinfo{person}{Aditi
  Dhabalia}, \bibinfo{person}{Hong Shen}, \bibinfo{person}{Kenneth Holstein},
  {and} \bibinfo{person}{Motahhare Eslami}.} \bibinfo{year}{2022}\natexlab{}.
\newblock \showarticletitle{Toward User-Driven Algorithm Auditing:
  Investigating Users’ Strategies for Uncovering Harmful Algorithmic
  Behavior}. In \bibinfo{booktitle}{\emph{Proceedings of the 2022 CHI
  Conference on Human Factors in Computing Systems}} (New Orleans, LA, USA)
  \emph{(\bibinfo{series}{CHI '22})}. \bibinfo{publisher}{Association for
  Computing Machinery}, \bibinfo{address}{New York, NY, USA}, Article
  \bibinfo{articleno}{626}, \bibinfo{numpages}{19}~pages.
\newblock
\showISBNx{9781450391573}
\urldef\tempurl%
\url{https://doi.org/10.1145/3491102.3517441}
\showDOI{\tempurl}


\bibitem[D\'{\i}az et~al\mbox{.}(2022)]%
        {diaz2022}
\bibfield{author}{\bibinfo{person}{Mark D\'{\i}az}, \bibinfo{person}{Ian
  Kivlichan}, \bibinfo{person}{Rachel Rosen}, \bibinfo{person}{Dylan Baker},
  \bibinfo{person}{Razvan Amironesei}, \bibinfo{person}{Vinodkumar
  Prabhakaran}, {and} \bibinfo{person}{Remi Denton}.}
  \bibinfo{year}{2022}\natexlab{}.
\newblock \showarticletitle{CrowdWorkSheets: Accounting for Individual and
  Collective Identities Underlying Crowdsourced Dataset Annotation}. In
  \bibinfo{booktitle}{\emph{2022 ACM Conference on Fairness, Accountability,
  and Transparency}} (Seoul, Republic of Korea) \emph{(\bibinfo{series}{FAccT
  '22})}. \bibinfo{publisher}{Association for Computing Machinery},
  \bibinfo{address}{New York, NY, USA}, \bibinfo{pages}{2342–2351}.
\newblock
\urldef\tempurl%
\url{https://doi.org/10.1145/3531146.3534647}
\showURL{%
\tempurl}


\bibitem[Doran(2016)]%
        {doran2016popular}
\bibfield{author}{\bibinfo{person}{Christine Doran}.}
  \bibinfo{year}{2016}\natexlab{}.
\newblock \showarticletitle{Popular Orientalism: Somerset Maugham in Mainland
  Southeast Asia}.
\newblock \bibinfo{journal}{\emph{Humanities}} \bibinfo{volume}{5},
  \bibinfo{number}{1} (\bibinfo{year}{2016}), \bibinfo{pages}{13}.
\newblock


\bibitem[Dourish(2016)]%
        {Dourish2016AlgorithmsAT}
\bibfield{author}{\bibinfo{person}{Paul Dourish}.}
  \bibinfo{year}{2016}\natexlab{}.
\newblock \showarticletitle{Algorithms and their others: Algorithmic culture in
  context}.
\newblock \bibinfo{journal}{\emph{Big Data \& Society}}  \bibinfo{volume}{3}
  (\bibinfo{year}{2016}).
\newblock


\bibitem[Eriksson~Krutr{\"o}k and {\AA}kerlund(2022)]%
        {eriksson2022through}
\bibfield{author}{\bibinfo{person}{Moa Eriksson~Krutr{\"o}k} {and}
  \bibinfo{person}{Mathilda {\AA}kerlund}.} \bibinfo{year}{2022}\natexlab{}.
\newblock \showarticletitle{Through a white lens: Black victimhood, visibility,
  and whiteness in the Black Lives Matter movement on TikTok}.
\newblock \bibinfo{journal}{\emph{Information, Communication \& Society}}
  (\bibinfo{year}{2022}), \bibinfo{pages}{1--19}.
\newblock


\bibitem[Fleras and Kunz(2001)]%
        {fleras2001media}
\bibfield{author}{\bibinfo{person}{Augie Fleras} {and}
  \bibinfo{person}{Jean~Lock Kunz}.} \bibinfo{year}{2001}\natexlab{}.
\newblock \bibinfo{booktitle}{\emph{Media and minorities: Representing
  diversity in a multicultural Canada}}.
\newblock \bibinfo{publisher}{Thompson Educational}.
\newblock


\bibitem[Gadiraju et~al\mbox{.}(2023)]%
        {Gadiraju2023}
\bibfield{author}{\bibinfo{person}{Vinitha Gadiraju}, \bibinfo{person}{Shaun
  Kane}, \bibinfo{person}{Sunipa Dev}, \bibinfo{person}{Alex Taylor},
  \bibinfo{person}{Ding Wang}, \bibinfo{person}{Remi Denton}, {and}
  \bibinfo{person}{Robin Brewer}.} \bibinfo{year}{2023}\natexlab{}.
\newblock \showarticletitle{"I wouldn't say offensive but...":
  Disability-Centered Perspectives on Large Language Models}. In
  \bibinfo{booktitle}{\emph{Proceedings of the 2023 Conference on Fairness,
  Accountability, and Transparency}}.
\newblock


\bibitem[Ge and Parikh(2021)]%
        {Ge_undated-sx}
\bibfield{author}{\bibinfo{person}{Songwei Ge} {and} \bibinfo{person}{Devi
  Parikh}.} \bibinfo{year}{2021}\natexlab{}.
\newblock \showarticletitle{Visual Conceptual Blending with Large-scale
  Language and Vision Models}.
\newblock \bibinfo{journal}{\emph{CoRR}}  \bibinfo{volume}{abs/2106.14127}
  (\bibinfo{year}{2021}).
\newblock
\showeprint[arXiv]{2106.14127}
\urldef\tempurl%
\url{https://arxiv.org/abs/2106.14127}
\showURL{%
\tempurl}


\bibitem[Gillespie(2016)]%
        {gillespie2016}
\bibfield{author}{\bibinfo{person}{Tarleton Gillespie}.}
  \bibinfo{year}{2016}\natexlab{}.
\newblock \bibinfo{booktitle}{\emph{\#Trendingistrending: When Algorithms
  Become Culture}}.
\newblock \bibinfo{publisher}{Routledge}.
\newblock
\urldef\tempurl%
\url{https://www.microsoft.com/en-us/research/publication/trendingistrending-when-algorithms-become-culture-3/}
\showURL{%
\tempurl}


\bibitem[Goyal et~al\mbox{.}(2022)]%
        {goyal2022}
\bibfield{author}{\bibinfo{person}{Nitesh Goyal}, \bibinfo{person}{Ian
  Kivlichan}, \bibinfo{person}{Rachel Rosen}, {and} \bibinfo{person}{Lucy
  Vasserman}.} \bibinfo{year}{2022}\natexlab{}.
\newblock \showarticletitle{Is Your Toxicity My Toxicity? Exploring the Impact
  of Rater Identity on Toxicity Annotation}.
\newblock \bibinfo{journal}{\emph{Proceedings of ACM Conference On
  Computer-Supported Cooperative Work And Social Computing (CSCW)}}
  (\bibinfo{year}{2022}).
\newblock


\bibitem[Griswold(2013)]%
        {Griswold2013}
\bibfield{author}{\bibinfo{person}{Wendy Griswold}.}
  \bibinfo{year}{2013}\natexlab{}.
\newblock \bibinfo{booktitle}{\emph{Cultures and Societies in a Changing
  World}}.
\newblock \bibinfo{publisher}{SAGE Publications, Inc.}
\newblock


\bibitem[Hall(1997)]%
        {hall1997representation}
\bibfield{author}{\bibinfo{person}{Stuart Hall}.}
  \bibinfo{year}{1997}\natexlab{}.
\newblock \bibinfo{booktitle}{\emph{Representation: cultural representations
  and signifying practices}}. Vol.~\bibinfo{volume}{1997}.
\newblock \bibinfo{publisher}{Sage London}.
\newblock


\bibitem[Hallinan and Striphas(2016)]%
        {Hallinan2016}
\bibfield{author}{\bibinfo{person}{Blake Hallinan} {and} \bibinfo{person}{Ted
  Striphas}.} \bibinfo{year}{2016}\natexlab{}.
\newblock \showarticletitle{Recommended for you: The Netflix Prize and the
  production of algorithmic culture}.
\newblock \bibinfo{journal}{\emph{New Media \& Society}} \bibinfo{volume}{18},
  \bibinfo{number}{1} (\bibinfo{year}{2016}), \bibinfo{pages}{117--137}.
\newblock
\urldef\tempurl%
\url{https://doi.org/10.1177/1461444814538646}
\showDOI{\tempurl}
\showeprint{https://doi.org/10.1177/1461444814538646}


\bibitem[Hanna et~al\mbox{.}(2020)]%
        {hanna2020}
\bibfield{author}{\bibinfo{person}{Alex Hanna}, \bibinfo{person}{Remi Denton},
  \bibinfo{person}{Andrew Smart}, {and} \bibinfo{person}{Jamila Smith-Loud}.}
  \bibinfo{year}{2020}\natexlab{}.
\newblock \showarticletitle{Towards a Critical Race Methodology in Algorithmic
  Fairness}. In \bibinfo{booktitle}{\emph{Proceedings of the 2020 Conference on
  Fairness, Accountability, and Transparency}} (Barcelona, Spain)
  \emph{(\bibinfo{series}{FAT* '20})}. \bibinfo{publisher}{Association for
  Computing Machinery}, \bibinfo{address}{New York, NY, USA},
  \bibinfo{pages}{501–512}.
\newblock
\showISBNx{9781450369367}
\urldef\tempurl%
\url{https://doi.org/10.1145/3351095.3372826}
\showDOI{\tempurl}


\bibitem[Harding(2013)]%
        {harding2013rethinking}
\bibfield{author}{\bibinfo{person}{Sandra Harding}.}
  \bibinfo{year}{2013}\natexlab{}.
\newblock \showarticletitle{Rethinking standpoint epistemology: What is
  “strong objectivity”?}
\newblock In \bibinfo{booktitle}{\emph{Feminist epistemologies}}.
  \bibinfo{publisher}{Routledge}, \bibinfo{pages}{49--82}.
\newblock


\bibitem[Hoffmann(2021)]%
        {hoffmann2021terms}
\bibfield{author}{\bibinfo{person}{Anna~Lauren Hoffmann}.}
  \bibinfo{year}{2021}\natexlab{}.
\newblock \showarticletitle{Terms of inclusion: Data, discourse, violence}.
\newblock \bibinfo{journal}{\emph{New Media \& Society}} \bibinfo{volume}{23},
  \bibinfo{number}{12} (\bibinfo{year}{2021}), \bibinfo{pages}{3539--3556}.
\newblock


\bibitem[hooks(1992)]%
        {hooks1992}
\bibfield{author}{\bibinfo{person}{bell hooks}.}
  \bibinfo{year}{1992}\natexlab{}.
\newblock \bibinfo{booktitle}{\emph{Black looks: Race and representation}}.
\newblock \bibinfo{publisher}{South End Press}.
\newblock


\bibitem[Hughes and DuMont(2002)]%
        {Hughes2002using}
\bibfield{author}{\bibinfo{person}{Diane~L Hughes} {and}
  \bibinfo{person}{Kimberly DuMont}.} \bibinfo{year}{2002}\natexlab{}.
\newblock \showarticletitle{Using focus groups to facilitate culturally
  anchored research}.
\newblock In \bibinfo{booktitle}{\emph{Ecological research to promote social
  change}}. \bibinfo{publisher}{Springer}, \bibinfo{pages}{257--289}.
\newblock


\bibitem[Hutchinson et~al\mbox{.}(2022)]%
        {hutchinson-2022-underspecification}
\bibfield{author}{\bibinfo{person}{Ben Hutchinson}, \bibinfo{person}{Jason
  Baldridge}, {and} \bibinfo{person}{Vinodkumar Prabhakaran}.}
  \bibinfo{year}{2022}\natexlab{}.
\newblock \showarticletitle{Underspecification in Scene
  Description-to-Depiction Tasks}. In \bibinfo{booktitle}{\emph{Proceedings of
  the 2nd Conference of the Asia-Pacific Chapter of the Association for
  Computational Linguistics and the 12th International Joint Conference on
  Natural Language Processing (Volume 1: Long Papers)}}.
  \bibinfo{publisher}{Association for Computational Linguistics},
  \bibinfo{address}{Online only}, \bibinfo{pages}{1172--1184}.
\newblock
\urldef\tempurl%
\url{https://aclanthology.org/2022.aacl-main.86}
\showURL{%
\tempurl}


\bibitem[Ibnazhar(2016)]%
        {heerranjha}
\bibfield{author}{\bibinfo{person}{Ibnazhar}.} \bibinfo{year}{2016}\natexlab{}.
\newblock \bibinfo{booktitle}{\emph{'Pakistan'- Islamabad Saidpur Village}}.
\newblock
\urldef\tempurl%
\url{https://www.google.com/url?q=https://commons.m.wikimedia.org/wiki/File:\%2527Pakistan\%2527-_Islamabad_Saidpur_Village_-_@ibneazhar_Sep_2016_(141).jpg&sa=D&source=docs&ust=1682549051045601&usg=AOvVaw3CYcKTtwuHxJ0gFyD4xuv_}
\showURL{%
\tempurl}


\bibitem[Inden(2001)]%
        {inden2001}
\bibfield{author}{\bibinfo{person}{Ronald Inden}.}
  \bibinfo{year}{2001}\natexlab{}.
\newblock \bibinfo{booktitle}{\emph{Imagining India}}.
\newblock \bibinfo{publisher}{Indiana University Press}.
\newblock


\bibitem[Jiwani(2009)]%
        {Jiwani2009}
\bibfield{author}{\bibinfo{person}{Yasmin Jiwani}.}
  \bibinfo{year}{2009}\natexlab{}.
\newblock \showarticletitle{Helpless Maidens and Chivalrous Knights: Afghan
  Women in the Canadian Press}.
\newblock \bibinfo{journal}{\emph{University of Toronto Quarterly}}
  \bibinfo{volume}{78}, \bibinfo{number}{2} (\bibinfo{year}{2009}),
  \bibinfo{pages}{728--744}.
\newblock
\urldef\tempurl%
\url{https://doi.org/10.3138/utq.78.2.728}
\showDOI{\tempurl}


\bibitem[Joseph(2002)]%
        {Joseph2002}
\bibfield{author}{\bibinfo{person}{Miranda Joseph}.}
  \bibinfo{year}{2002}\natexlab{}.
\newblock \bibinfo{booktitle}{\emph{Against the Romance of Community}}.
\newblock \bibinfo{publisher}{University of Minnesota Press}.
\newblock


\bibitem[Kak(2020)]%
        {kak2020}
\bibfield{author}{\bibinfo{person}{Amba Kak}.} \bibinfo{year}{2020}\natexlab{}.
\newblock \showarticletitle{"The Global South is Everywhere, but Also Always
  Somewhere": National Policy Narratives and AI Justice}. In
  \bibinfo{booktitle}{\emph{Proceedings of the AAAI/ACM Conference on AI,
  Ethics, and Society}} (New York, NY, USA) \emph{(\bibinfo{series}{AIES
  '20})}. \bibinfo{publisher}{Association for Computing Machinery},
  \bibinfo{address}{New York, NY, USA}, \bibinfo{pages}{307–312}.
\newblock
\showISBNx{9781450371100}
\urldef\tempurl%
\url{https://doi.org/10.1145/3375627.3375859}
\showDOI{\tempurl}


\bibitem[Kalluri(2020)]%
        {kalluri2020}
\bibfield{author}{\bibinfo{person}{Pratyusha Kalluri}.}
  \bibinfo{year}{2020}\natexlab{}.
\newblock \bibinfo{title}{Don’t ask if artificial intelligence is good or
  fair, ask how it shifts power}.
\newblock \bibinfo{howpublished}{Nature}.
\newblock
\urldef\tempurl%
\url{https://www.nature.com/articles/d41586-020-02003-2}
\showURL{%
\tempurl}


\bibitem[Kapania et~al\mbox{.}(2022)]%
        {kapania2022}
\bibfield{author}{\bibinfo{person}{Shivani Kapania}, \bibinfo{person}{Oliver
  Siy}, \bibinfo{person}{Gabe Clapper}, \bibinfo{person}{Azhagu~Meena SP},
  {and} \bibinfo{person}{Nithya Sambasivan}.} \bibinfo{year}{2022}\natexlab{}.
\newblock \showarticletitle{”Because AI is 100\% Right and Safe”: User
  Attitudes and Sources of AI Authority in India}. In
  \bibinfo{booktitle}{\emph{Proceedings of the 2022 CHI Conference on Human
  Factors in Computing Systems}} (New Orleans, LA, USA)
  \emph{(\bibinfo{series}{CHI '22})}. \bibinfo{publisher}{Association for
  Computing Machinery}, \bibinfo{address}{New York, NY, USA}, Article
  \bibinfo{articleno}{158}, \bibinfo{numpages}{18}~pages.
\newblock
\showISBNx{9781450391573}
\urldef\tempurl%
\url{https://doi.org/10.1145/3491102.3517533}
\showDOI{\tempurl}


\bibitem[Karim(1997)]%
        {Karim1997}
\bibfield{author}{\bibinfo{person}{Karim~H. Karim}.}
  \bibinfo{year}{1997}\natexlab{}.
\newblock \showarticletitle{The historical resilience of primary stereotypes:
  Core images of the Muslim Other}.
\newblock In \bibinfo{booktitle}{\emph{The language and politics of exclusion:
  Others in discourse}}, \bibfield{editor}{\bibinfo{person}{Stephen~H.
  Riggins}} (Ed.). \bibinfo{publisher}{Thousand Oaks, CA: Sage Publications},
  \bibinfo{pages}{153--182}.
\newblock


\bibitem[Karizat et~al\mbox{.}(2021)]%
        {karizat2021}
\bibfield{author}{\bibinfo{person}{Nadia Karizat}, \bibinfo{person}{Dan
  Delmonaco}, \bibinfo{person}{Motahhare Eslami}, {and}
  \bibinfo{person}{Nazanin Andalibi}.} \bibinfo{year}{2021}\natexlab{}.
\newblock \showarticletitle{Algorithmic Folk Theories and Identity: How TikTok
  Users Co-Produce Knowledge of Identity and Engage in Algorithmic Resistance}.
\newblock \bibinfo{journal}{\emph{Proc. ACM Hum.-Comput. Interact.}}
  \bibinfo{volume}{5}, \bibinfo{number}{CSCW2} (\bibinfo{year}{2021}).
\newblock


\bibitem[Khan(2017)]%
        {Baitul}
\bibfield{author}{\bibinfo{person}{Md. Nazmul~Hasan Khan}.}
  \bibinfo{year}{2017}\natexlab{}.
\newblock \bibinfo{booktitle}{\emph{Baitul Mukarram National Mosque, Dhaka,
  Bangladesh}}.
\newblock
\urldef\tempurl%
\url{https://upload.wikimedia.org/wikipedia/commons/6/60/Baitul_Mukarram_National_Mosque\%2C_Dhaka\%2C_Bangladesh.jpg}
\showURL{%
\tempurl}


\bibitem[Kirk et~al\mbox{.}(1986)]%
        {kirk1986reliability}
\bibfield{author}{\bibinfo{person}{Jerome Kirk}, \bibinfo{person}{Marc~L
  Miller}, {and} \bibinfo{person}{Marc~Louis Miller}.}
  \bibinfo{year}{1986}\natexlab{}.
\newblock \bibinfo{booktitle}{\emph{Reliability and validity in qualitative
  research}}.
\newblock \bibinfo{publisher}{Sage}.
\newblock


\bibitem[Kothari(2001)]%
        {Kothari2001}
\bibfield{author}{\bibinfo{person}{U. Kothari}.}
  \bibinfo{year}{2001}\natexlab{}.
\newblock \showarticletitle{Power, knowledge and social control in
  participatory development}.
\newblock In \bibinfo{booktitle}{\emph{Participation: the new tyranny?}},
  \bibfield{editor}{\bibinfo{person}{William Cooke} {and}
  \bibinfo{person}{U.~Kothari}} (Eds.). \bibinfo{publisher}{Zed Books},
  \bibinfo{address}{United Kingdom}, \bibinfo{pages}{139--152}.
\newblock


\bibitem[Krueger and Casey(2000)]%
        {krueger2000practical}
\bibfield{author}{\bibinfo{person}{Richard~A Krueger} {and}
  \bibinfo{person}{Mary~Anne Casey}.} \bibinfo{year}{2000}\natexlab{}.
\newblock \bibinfo{booktitle}{\emph{Focus Groups: A Practical Guide for Applied
  Research}}.
\newblock \bibinfo{publisher}{SAGE Publications}, \bibinfo{address}{Thousand
  Oaks, CA}.
\newblock


\bibitem[Lally(2019)]%
        {lally2019colour}
\bibfield{author}{\bibinfo{person}{Jagjeet Lally}.}
  \bibinfo{year}{2019}\natexlab{}.
\newblock \showarticletitle{Colour as Commodity: Colonialism and the sensory
  worlds of South Asia}. In \bibinfo{booktitle}{\emph{Third Text Forum
  online}}. Taylor \& Francis.
\newblock


\bibitem[Lau(2009)]%
        {lau2009}
\bibfield{author}{\bibinfo{person}{Lisa Lau}.} \bibinfo{year}{2009}\natexlab{}.
\newblock \bibinfo{booktitle}{\emph{Re-Orientalism: The Perpetration and
  Development of Orientalism by Orientals}}.
\newblock \bibinfo{publisher}{Cambridge University Press}.
\newblock


\bibitem[Lau and Mendes(2011)]%
        {lau2014}
\bibfield{author}{\bibinfo{person}{Lisa Lau} {and}
  \bibinfo{person}{Ana~Cristina Mendes}.} \bibinfo{year}{2011}\natexlab{}.
\newblock \showarticletitle{Introducing re-Orientalism: A new manifestation of
  Orientalism}.
\newblock In \bibinfo{booktitle}{\emph{Re-Orientalism and South Asian Identity
  Politics: The Oriental Other Within}}. \bibinfo{publisher}{Routledge},
  \bibinfo{pages}{1--14}.
\newblock


\bibitem[Lewis(2004)]%
        {lewis2004rethinking}
\bibfield{author}{\bibinfo{person}{Reina Lewis}.}
  \bibinfo{year}{2004}\natexlab{}.
\newblock \bibinfo{booktitle}{\emph{Rethinking Orientalism: Women, travel and
  the Ottoman harem}}. Vol.~\bibinfo{volume}{4}.
\newblock \bibinfo{publisher}{Taylor \& Francis}.
\newblock


\bibitem[Maira(2008)]%
        {maira2008belly}
\bibfield{author}{\bibinfo{person}{Sunaina Maira}.}
  \bibinfo{year}{2008}\natexlab{}.
\newblock \showarticletitle{Belly dancing: Arab-face, Orientalist feminism, and
  US empire}.
\newblock \bibinfo{journal}{\emph{American Quarterly}} \bibinfo{volume}{60},
  \bibinfo{number}{2} (\bibinfo{year}{2008}), \bibinfo{pages}{317--345}.
\newblock


\bibitem[Malreddy(2015)]%
        {malreddy2015}
\bibfield{author}{\bibinfo{person}{Pavan~Kumar Malreddy}.}
  \bibinfo{year}{2015}\natexlab{}.
\newblock \bibinfo{booktitle}{\emph{Orientalism, Terrorism, Indigenism: South
  Asian Readings in Postcolonialism}}.
\newblock \bibinfo{publisher}{Sage Publications}.
\newblock


\bibitem[McKinnon(2006)]%
        {McKinnon2006}
\bibfield{author}{\bibinfo{person}{K McKinnon}.}
  \bibinfo{year}{2006}\natexlab{}.
\newblock \showarticletitle{An orthodoxy of 'the local': post-colonialism,
  participation and professionalism in northern Thailand}.
\newblock \bibinfo{journal}{\emph{The Geographical Journal}}
  \bibinfo{volume}{172} (\bibinfo{year}{2006}), \bibinfo{pages}{22--34}.
\newblock
\urldef\tempurl%
\url{https://doi.org/10.1111/j.1475-4959.2006.00182.x}
\showDOI{\tempurl}


\bibitem[Mengesha et~al\mbox{.}(2021)]%
        {mengesha2021don}
\bibfield{author}{\bibinfo{person}{Zion Mengesha}, \bibinfo{person}{Courtney
  Heldreth}, \bibinfo{person}{Michal Lahav}, \bibinfo{person}{Juliana
  Sublewski}, {and} \bibinfo{person}{Elyse Tuennerman}.}
  \bibinfo{year}{2021}\natexlab{}.
\newblock \showarticletitle{“I Don’t Think These Devices are Very
  Culturally Sensitive.”—Impact of Automated Speech Recognition Errors on
  African Americans}.
\newblock \bibinfo{journal}{\emph{Frontiers in Artificial Intelligence}}
  (\bibinfo{year}{2021}), \bibinfo{pages}{169}.
\newblock


\bibitem[Mohamed et~al\mbox{.}(2020)]%
        {mohamed2020decolonial}
\bibfield{author}{\bibinfo{person}{Shakir Mohamed},
  \bibinfo{person}{Marie-Therese Png}, {and} \bibinfo{person}{William Isaac}.}
  \bibinfo{year}{2020}\natexlab{}.
\newblock \showarticletitle{Decolonial AI: Decolonial theory as sociotechnical
  foresight in artificial intelligence}.
\newblock \bibinfo{journal}{\emph{Philosophy \& Technology}}
  \bibinfo{volume}{33} (\bibinfo{year}{2020}), \bibinfo{pages}{659--684}.
\newblock


\bibitem[Mohanty(1988)]%
        {Mohanty88}
\bibfield{author}{\bibinfo{person}{Chandra Mohanty}.}
  \bibinfo{year}{1988}\natexlab{}.
\newblock \showarticletitle{Under Western Eyes: Feminist Scholarship and
  Colonial Discourses}.
\newblock \bibinfo{journal}{\emph{Feminist Review}} \bibinfo{volume}{30},
  \bibinfo{number}{1} (\bibinfo{year}{1988}), \bibinfo{pages}{61--88}.
\newblock
\urldef\tempurl%
\url{https://doi.org/10.1057/fr.1988.42}
\showDOI{\tempurl}
\showeprint{https://doi.org/10.1057/fr.1988.42}


\bibitem[Nacos and Torres-Reyna(2004)]%
        {nacos2004framing}
\bibfield{author}{\bibinfo{person}{Brigitte~L Nacos} {and}
  \bibinfo{person}{Oscar Torres-Reyna}.} \bibinfo{year}{2004}\natexlab{}.
\newblock \showarticletitle{Framing Muslim-Americans before and after 9/11}.
\newblock In \bibinfo{booktitle}{\emph{Framing Terrorism}}.
  \bibinfo{publisher}{Routledge}, \bibinfo{pages}{141--166}.
\newblock


\bibitem[Nakamura(2007)]%
        {nakamura2008}
\bibfield{author}{\bibinfo{person}{Lisa Nakamura}.}
  \bibinfo{year}{2007}\natexlab{}.
\newblock \bibinfo{booktitle}{\emph{Digitizing race: Visual cultures of the
  internet}}.
\newblock \bibinfo{publisher}{University of Minnesota Press}.
\newblock


\bibitem[Oh(2022)]%
        {oh2022whitewashing}
\bibfield{author}{\bibinfo{person}{David Oh}.} \bibinfo{year}{2022}\natexlab{}.
\newblock \bibinfo{title}{Whitewashing the Movies: Asian Erasure and White
  Subjectivity in US Film Culture}.
\newblock
\newblock


\bibitem[Onwuegbuzie et~al\mbox{.}(2009)]%
        {onwuegbuzie2009qualitative}
\bibfield{author}{\bibinfo{person}{Anthony~J Onwuegbuzie},
  \bibinfo{person}{Wendy~B Dickinson}, \bibinfo{person}{Nancy~L Leech}, {and}
  \bibinfo{person}{Annmarie~G Zoran}.} \bibinfo{year}{2009}\natexlab{}.
\newblock \showarticletitle{A Qualitative Framework for Collecting and
  Analyzing Data in Focus Group Research}.
\newblock \bibinfo{journal}{\emph{International Journal of Qualitative
  Methods}} \bibinfo{volume}{8}, \bibinfo{number}{3} (\bibinfo{date}{Sept.}
  \bibinfo{year}{2009}), \bibinfo{pages}{1--21}.
\newblock


\bibitem[Palinkas et~al\mbox{.}(2015)]%
        {palinkas2015purposeful}
\bibfield{author}{\bibinfo{person}{Lawrence~A Palinkas},
  \bibinfo{person}{Sarah~M Horwitz}, \bibinfo{person}{Carla~A Green},
  \bibinfo{person}{Jennifer~P Wisdom}, \bibinfo{person}{Naihua Duan}, {and}
  \bibinfo{person}{Kimberly Hoagwood}.} \bibinfo{year}{2015}\natexlab{}.
\newblock \showarticletitle{Purposeful sampling for qualitative data collection
  and analysis in mixed method implementation research}.
\newblock \bibinfo{journal}{\emph{Administration and policy in mental health
  and mental health services research}}  \bibinfo{volume}{42}
  (\bibinfo{year}{2015}), \bibinfo{pages}{533--544}.
\newblock


\bibitem[Paullada et~al\mbox{.}(2021)]%
        {paullada2021data}
\bibfield{author}{\bibinfo{person}{Amandalynne Paullada},
  \bibinfo{person}{Inioluwa~Deborah Raji}, \bibinfo{person}{Emily~M Bender},
  \bibinfo{person}{Remi Denton}, {and} \bibinfo{person}{Alex Hanna}.}
  \bibinfo{year}{2021}\natexlab{}.
\newblock \showarticletitle{Data and its (dis) contents: A survey of dataset
  development and use in machine learning research}.
\newblock \bibinfo{journal}{\emph{Patterns}} \bibinfo{volume}{2},
  \bibinfo{number}{11} (\bibinfo{year}{2021}), \bibinfo{pages}{100336}.
\newblock


\bibitem[Png(2022)]%
        {png2022}
\bibfield{author}{\bibinfo{person}{Marie-Therese Png}.}
  \bibinfo{year}{2022}\natexlab{}.
\newblock \showarticletitle{At the Tensions of South and North: Critical Roles
  of Global South Stakeholders in AI Governance}. In
  \bibinfo{booktitle}{\emph{2022 ACM Conference on Fairness, Accountability,
  and Transparency}} (Seoul, Republic of Korea) \emph{(\bibinfo{series}{FAccT
  '22})}. \bibinfo{publisher}{Association for Computing Machinery},
  \bibinfo{address}{New York, NY, USA}, \bibinfo{pages}{1434–1445}.
\newblock
\showISBNx{9781450393522}
\urldef\tempurl%
\url{https://doi.org/10.1145/3531146.3533200}
\showDOI{\tempurl}


\bibitem[Prabhakaran and Martin(2020)]%
        {Prabhakaran2020}
\bibfield{author}{\bibinfo{person}{Vinodkumar Prabhakaran} {and}
  \bibinfo{person}{Donald Martin}.} \bibinfo{year}{2020}\natexlab{}.
\newblock \showarticletitle{Participatory Machine Learning using Community
  Based System Dynamics}.
\newblock \bibinfo{journal}{\emph{Health and Human Rights Journal}}
  (\bibinfo{year}{2020}).
\newblock


\bibitem[Prabhakaran et~al\mbox{.}(2022)]%
        {prabhakaran2022cultural}
\bibfield{author}{\bibinfo{person}{Vinodkumar Prabhakaran},
  \bibinfo{person}{Rida Qadri}, {and} \bibinfo{person}{Ben Hutchinson}.}
  \bibinfo{year}{2022}\natexlab{}.
\newblock \showarticletitle{Cultural Incongruencies in Artificial
  Intelligence}.
\newblock \bibinfo{journal}{\emph{arXiv preprint arXiv:2211.13069}}
  (\bibinfo{year}{2022}).
\newblock


\bibitem[Radford et~al\mbox{.}(2021)]%
        {radford2021}
\bibfield{author}{\bibinfo{person}{Alec Radford}, \bibinfo{person}{Jong~Wook
  Kim}, \bibinfo{person}{Chris Hallacy}, \bibinfo{person}{Aditya Ramesh},
  \bibinfo{person}{Gabriel Goh}, \bibinfo{person}{Sandhini Agarwal},
  \bibinfo{person}{Girish Sastry}, \bibinfo{person}{Amanda Askell},
  \bibinfo{person}{Pamela Mishkin}, \bibinfo{person}{Jack Clark},
  \bibinfo{person}{Gretchen Krueger}, {and} \bibinfo{person}{Ilya Sutskever}.}
  \bibinfo{year}{2021}\natexlab{}.
\newblock \showarticletitle{Learning Transferable Visual Models From Natural
  Language Supervision}. In \bibinfo{booktitle}{\emph{Proceedings of the 38th
  International Conference on Machine Learning}}
  \emph{(\bibinfo{series}{Proceedings of Machine Learning Research},
  Vol.~\bibinfo{volume}{139})}, \bibfield{editor}{\bibinfo{person}{Marina
  Meila} {and} \bibinfo{person}{Tong Zhang}} (Eds.). \bibinfo{publisher}{PMLR},
  \bibinfo{pages}{8748--8763}.
\newblock
\urldef\tempurl%
\url{https://proceedings.mlr.press/v139/radford21a.html}
\showURL{%
\tempurl}


\bibitem[Ramesh et~al\mbox{.}(2022)]%
        {dalle}
\bibfield{author}{\bibinfo{person}{Aditya Ramesh}, \bibinfo{person}{Prafulla
  Dhariwal}, \bibinfo{person}{Alex Nichol}, \bibinfo{person}{Casey Chu}, {and}
  \bibinfo{person}{Mark Chen}.} \bibinfo{year}{2022}\natexlab{}.
\newblock \showarticletitle{Hierarchical Text-Conditional Image Generation with
  {CLIP} Latents}.
\newblock \bibinfo{journal}{\emph{CoRR}}  \bibinfo{volume}{abs/2204.06125}
  (\bibinfo{year}{2022}).
\newblock
\urldef\tempurl%
\url{https://doi.org/10.48550/arXiv.2204.06125}
\showDOI{\tempurl}
\showeprint[arXiv]{2204.06125}


\bibitem[Ramesh et~al\mbox{.}(2021)]%
        {ramesh2021}
\bibfield{author}{\bibinfo{person}{Aditya Ramesh}, \bibinfo{person}{Mikhail
  Pavlov}, \bibinfo{person}{Gabriel Goh}, \bibinfo{person}{Scott Gray},
  \bibinfo{person}{Chelsea Voss}, \bibinfo{person}{Alec Radford},
  \bibinfo{person}{Mark Chen}, {and} \bibinfo{person}{Ilya Sutskever}.}
  \bibinfo{year}{2021}\natexlab{}.
\newblock \showarticletitle{Zero-Shot Text-to-Image Generation}. In
  \bibinfo{booktitle}{\emph{Proceedings of the 38th International Conference on
  Machine Learning}} \emph{(\bibinfo{series}{Proceedings of Machine Learning
  Research}, Vol.~\bibinfo{volume}{139})},
  \bibfield{editor}{\bibinfo{person}{Marina Meila} {and} \bibinfo{person}{Tong
  Zhang}} (Eds.). \bibinfo{publisher}{PMLR}, \bibinfo{pages}{8821--8831}.
\newblock
\urldef\tempurl%
\url{https://proceedings.mlr.press/v139/ramesh21a.html}
\showURL{%
\tempurl}


\bibitem[Rao(2009)]%
        {rao2009caste}
\bibfield{author}{\bibinfo{person}{Anupama Rao}.}
  \bibinfo{year}{2009}\natexlab{}.
\newblock \bibinfo{booktitle}{\emph{The caste question: Dalits and the politics
  of modern India}}.
\newblock \bibinfo{publisher}{Univ of California Press}.
\newblock


\bibitem[Rodriguez et~al\mbox{.}(2011)]%
        {rodriguez2011culturally}
\bibfield{author}{\bibinfo{person}{Katrina~L Rodriguez},
  \bibinfo{person}{Jana~L Schwartz}, \bibinfo{person}{Maria~KE Lahman}, {and}
  \bibinfo{person}{Monica~R Geist}.} \bibinfo{year}{2011}\natexlab{}.
\newblock \showarticletitle{Culturally responsive focus groups: Reframing the
  research experience to focus on participants}.
\newblock \bibinfo{journal}{\emph{International Journal of Qualitative
  Methods}} \bibinfo{volume}{10}, \bibinfo{number}{4} (\bibinfo{year}{2011}),
  \bibinfo{pages}{400--417}.
\newblock


\bibitem[Rombach et~al\mbox{.}(2022)]%
        {stable-diffusion}
\bibfield{author}{\bibinfo{person}{Robin Rombach}, \bibinfo{person}{Andreas
  Blattmann}, \bibinfo{person}{Dominik Lorenz}, \bibinfo{person}{Patrick
  Esser}, {and} \bibinfo{person}{BjÃ¶rn Ommer}.}
  \bibinfo{year}{2022}\natexlab{}.
\newblock \showarticletitle{High-Resolution Image Synthesis with Latent
  Diffusion Models}. In \bibinfo{booktitle}{\emph{Proceedings of the IEEE
  Conference on Computer Vision and Pattern Recognition (CVPR)}}.
\newblock
\urldef\tempurl%
\url{https://github.com/CompVis/latent-diffusionhttps://arxiv.org/abs/2112.10752}
\showURL{%
\tempurl}


\bibitem[Rose(1983)]%
        {rose1983hand}
\bibfield{author}{\bibinfo{person}{Hilary Rose}.}
  \bibinfo{year}{1983}\natexlab{}.
\newblock \showarticletitle{Hand, brain, and heart: A feminist epistemology for
  the natural sciences}.
\newblock \bibinfo{journal}{\emph{Signs: journal of Women in Culture and
  Society}} \bibinfo{volume}{9}, \bibinfo{number}{1} (\bibinfo{year}{1983}),
  \bibinfo{pages}{73--90}.
\newblock


\bibitem[Rubinstein and Sluis(2013)]%
        {rubinstein2013digital}
\bibfield{author}{\bibinfo{person}{Daniel Rubinstein} {and}
  \bibinfo{person}{Katrina Sluis}.} \bibinfo{year}{2013}\natexlab{}.
\newblock \showarticletitle{The digital image in photographic culture:
  Algorithmic photography and the crisis of representation}.
\newblock In \bibinfo{booktitle}{\emph{The photographic image in digital
  culture}}. \bibinfo{publisher}{Routledge}, \bibinfo{pages}{22--40}.
\newblock


\bibitem[Sadequain and Siddiqui(2012)]%
        {Sadequain}
\bibfield{author}{\bibinfo{person}{Sadequain} {and}
  \bibinfo{person}{Madiha~Naqsh Siddiqui}.} \bibinfo{year}{2012}\natexlab{}.
\newblock \bibinfo{booktitle}{\emph{Ceiling of Frere Hall}}.
\newblock
\urldef\tempurl%
\url{https://www.google.com/url?q=https://commons.wikimedia.org/wiki/Category:Painted_ceiling_of_Frere_Hall\%23/media/File:Ceiling_of_Frere_Hall_(2).jpg&sa=D&source=docs&ust=1675641953094542&usg=AOvVaw3HmquTXox7RYu_-DYcdIUx}
\showURL{%
\tempurl}


\bibitem[Sadowski and Selinger(2014)]%
        {sadowski2014creating}
\bibfield{author}{\bibinfo{person}{Jathan Sadowski} {and} \bibinfo{person}{Evan
  Selinger}.} \bibinfo{year}{2014}\natexlab{}.
\newblock \showarticletitle{Creating a taxonomic tool for technocracy and
  applying it to Silicon Valley}.
\newblock \bibinfo{journal}{\emph{Technology in Society}}  \bibinfo{volume}{38}
  (\bibinfo{year}{2014}), \bibinfo{pages}{161--168}.
\newblock


\bibitem[Saharia et~al\mbox{.}(2022a)]%
        {saharia2022photorealistic}
\bibfield{author}{\bibinfo{person}{Chitwan Saharia}, \bibinfo{person}{William
  Chan}, \bibinfo{person}{Saurabh Saxena}, \bibinfo{person}{Lala Li},
  \bibinfo{person}{Jay Whang}, \bibinfo{person}{Remi Denton},
  \bibinfo{person}{Seyed Kamyar~Seyed Ghasemipour},
  \bibinfo{person}{Burcu~Karagol Ayan}, \bibinfo{person}{S~Sara Mahdavi},
  \bibinfo{person}{Rapha~Gontijo Lopes}, {et~al\mbox{.}}}
  \bibinfo{year}{2022}\natexlab{a}.
\newblock \showarticletitle{Photorealistic Text-to-Image Diffusion Models with
  Deep Language Understanding}.
\newblock \bibinfo{journal}{\emph{arXiv preprint arXiv:2205.11487}}
  (\bibinfo{year}{2022}).
\newblock


\bibitem[Saharia et~al\mbox{.}(2022b)]%
        {imagen}
\bibfield{author}{\bibinfo{person}{Chitwan Saharia}, \bibinfo{person}{William
  Chan}, \bibinfo{person}{Saurabh Saxena}, \bibinfo{person}{Lala Li},
  \bibinfo{person}{Jay Whang}, \bibinfo{person}{Remi Denton},
  \bibinfo{person}{Seyed Kamyar~Seyed Ghasemipour},
  \bibinfo{person}{Burcu~Karagol Ayan}, \bibinfo{person}{S~Sara Mahdavi},
  \bibinfo{person}{Rapha~Gontijo Lopes}, {et~al\mbox{.}}}
  \bibinfo{year}{2022}\natexlab{b}.
\newblock \showarticletitle{Photorealistic text-to-image diffusion models with
  deep language understanding}.
\newblock \bibinfo{journal}{\emph{Advances in Neural Information Processing
  Systems}} (\bibinfo{year}{2022}).
\newblock


\bibitem[Said(1978)]%
        {said1978}
\bibfield{author}{\bibinfo{person}{Edward~W. Said}.}
  \bibinfo{year}{1978}\natexlab{}.
\newblock \bibinfo{booktitle}{\emph{Orientalism}}.
\newblock \bibinfo{publisher}{Pantheon Books}.
\newblock


\bibitem[Sambasivan(2022)]%
        {sambasivan2022}
\bibfield{author}{\bibinfo{person}{Nithya Sambasivan}.}
  \bibinfo{year}{2022}\natexlab{}.
\newblock \showarticletitle{All Equation, No Human: The Myopia of AI Models}.
\newblock \bibinfo{journal}{\emph{Interactions}} \bibinfo{volume}{29},
  \bibinfo{number}{2} (\bibinfo{date}{feb} \bibinfo{year}{2022}),
  \bibinfo{pages}{78–80}.
\newblock
\showISSN{1072-5520}
\urldef\tempurl%
\url{https://doi.org/10.1145/3516515}
\showDOI{\tempurl}


\bibitem[Sambasivan et~al\mbox{.}(2021a)]%
        {sambasivan2021}
\bibfield{author}{\bibinfo{person}{Nithya Sambasivan}, \bibinfo{person}{Erin
  Arnesen}, \bibinfo{person}{Ben Hutchinson}, \bibinfo{person}{Tulsee Doshi},
  {and} \bibinfo{person}{Vinodkumar Prabhakaran}.}
  \bibinfo{year}{2021}\natexlab{a}.
\newblock \showarticletitle{Re-Imagining Algorithmic Fairness in India and
  Beyond}. In \bibinfo{booktitle}{\emph{Proceedings of the 2021 ACM Conference
  on Fairness, Accountability, and Transparency}} (Virtual Event, Canada)
  \emph{(\bibinfo{series}{FAccT '21})}. \bibinfo{publisher}{Association for
  Computing Machinery}, \bibinfo{address}{New York, NY, USA},
  \bibinfo{pages}{315–328}.
\newblock
\showISBNx{9781450383097}
\urldef\tempurl%
\url{https://doi.org/10.1145/3442188.3445896}
\showDOI{\tempurl}


\bibitem[Sambasivan et~al\mbox{.}(2021b)]%
        {sambasivan2021re}
\bibfield{author}{\bibinfo{person}{Nithya Sambasivan}, \bibinfo{person}{Erin
  Arnesen}, \bibinfo{person}{Ben Hutchinson}, \bibinfo{person}{Tulsee Doshi},
  {and} \bibinfo{person}{Vinodkumar Prabhakaran}.}
  \bibinfo{year}{2021}\natexlab{b}.
\newblock \showarticletitle{Re-imagining algorithmic fairness in india and
  beyond}. In \bibinfo{booktitle}{\emph{Proceedings of the 2021 ACM conference
  on fairness, accountability, and transparency}}. \bibinfo{pages}{315--328}.
\newblock


\bibitem[Sambasivan et~al\mbox{.}(2020)]%
        {sambasivan2020non}
\bibfield{author}{\bibinfo{person}{Nithya Sambasivan}, \bibinfo{person}{Erin
  Arnesen}, \bibinfo{person}{Ben Hutchinson}, {and} \bibinfo{person}{Vinodkumar
  Prabhakaran}.} \bibinfo{year}{2020}\natexlab{}.
\newblock \showarticletitle{Non-portability of Algorithmic Fairness in India}.
\newblock \bibinfo{journal}{\emph{arXiv preprint arXiv:2012.03659}}
  (\bibinfo{year}{2020}).
\newblock


\bibitem[Seaver(2017)]%
        {Seaver2017}
\bibfield{author}{\bibinfo{person}{Nick Seaver}.}
  \bibinfo{year}{2017}\natexlab{}.
\newblock \showarticletitle{Algorithms as culture: Some tactics for the
  ethnography of algorithmic systems}.
\newblock \bibinfo{journal}{\emph{Big Data \& Society}} \bibinfo{volume}{4},
  \bibinfo{number}{2} (\bibinfo{year}{2017}),
  \bibinfo{pages}{2053951717738104}.
\newblock
\urldef\tempurl%
\url{https://doi.org/10.1177/2053951717738104}
\showDOI{\tempurl}
\showeprint{https://doi.org/10.1177/2053951717738104}


\bibitem[Selbst et~al\mbox{.}(2019)]%
        {selbst2019fairness}
\bibfield{author}{\bibinfo{person}{Andrew~D Selbst}, \bibinfo{person}{Danah
  Boyd}, \bibinfo{person}{Sorelle~A Friedler}, \bibinfo{person}{Suresh
  Venkatasubramanian}, {and} \bibinfo{person}{Janet Vertesi}.}
  \bibinfo{year}{2019}\natexlab{}.
\newblock \showarticletitle{Fairness and abstraction in sociotechnical
  systems}. In \bibinfo{booktitle}{\emph{Proceedings of the conference on
  fairness, accountability, and transparency}}. \bibinfo{pages}{59--68}.
\newblock


\bibitem[Shelby et~al\mbox{.}(2022)]%
        {shelby2022sociotechnical}
\bibfield{author}{\bibinfo{person}{Renee Shelby}, \bibinfo{person}{Shalaleh
  Rismani}, \bibinfo{person}{Kathryn Henne}, \bibinfo{person}{AJung Moon},
  \bibinfo{person}{Negar Rostamzadeh}, \bibinfo{person}{Paul Nicholas},
  \bibinfo{person}{N'Mah Yilla}, \bibinfo{person}{Jess Gallegos},
  \bibinfo{person}{Andrew Smart}, \bibinfo{person}{Emilio Garcia},
  {et~al\mbox{.}}} \bibinfo{year}{2022}\natexlab{}.
\newblock \showarticletitle{Sociotechnical Harms: Scoping a Taxonomy for Harm
  Reduction}.
\newblock \bibinfo{journal}{\emph{arXiv preprint arXiv:2210.05791}}
  (\bibinfo{year}{2022}).
\newblock


\bibitem[Shen et~al\mbox{.}(2022)]%
        {shen2022}
\bibfield{author}{\bibinfo{person}{Hong Shen}, \bibinfo{person}{Leijie Wang},
  \bibinfo{person}{Wesley~H. Deng}, \bibinfo{person}{Ciell Brusse},
  \bibinfo{person}{Ronald Velgersdijk}, {and} \bibinfo{person}{Haiyi Zhu}.}
  \bibinfo{year}{2022}\natexlab{}.
\newblock \showarticletitle{The Model Card Authoring Toolkit: Toward
  Community-Centered, Deliberation-Driven AI Design}. In
  \bibinfo{booktitle}{\emph{2022 ACM Conference on Fairness, Accountability,
  and Transparency}} (Seoul, Republic of Korea) \emph{(\bibinfo{series}{FAccT
  '22})}. \bibinfo{publisher}{Association for Computing Machinery},
  \bibinfo{address}{New York, NY, USA}, \bibinfo{pages}{440–451}.
\newblock
\showISBNx{9781450393522}
\urldef\tempurl%
\url{https://doi.org/10.1145/3531146.3533110}
\showDOI{\tempurl}


\bibitem[Sloane et~al\mbox{.}(2022)]%
        {sloane2022}
\bibfield{author}{\bibinfo{person}{Mona Sloane}, \bibinfo{person}{Emanuel
  Moss}, \bibinfo{person}{Olaitan Awomolo}, {and} \bibinfo{person}{Laura
  Forlano}.} \bibinfo{year}{2022}\natexlab{}.
\newblock \showarticletitle{Participation Is Not a Design Fix for Machine
  Learning}. In \bibinfo{booktitle}{\emph{Equity and Access in Algorithms,
  Mechanisms, and Optimization}} (Arlington, VA, USA)
  \emph{(\bibinfo{series}{EAAMO '22})}. \bibinfo{publisher}{Association for
  Computing Machinery}, \bibinfo{address}{New York, NY, USA}, Article
  \bibinfo{articleno}{1}, \bibinfo{numpages}{6}~pages.
\newblock
\showISBNx{9781450394772}
\urldef\tempurl%
\url{https://doi.org/10.1145/3551624.3555285}
\showDOI{\tempurl}


\bibitem[Smith(1987)]%
        {smith1987everyday}
\bibfield{author}{\bibinfo{person}{Dorothy~E Smith}.}
  \bibinfo{year}{1987}\natexlab{}.
\newblock \bibinfo{booktitle}{\emph{The everyday world as problematic: A
  feminist sociology}}.
\newblock \bibinfo{publisher}{University of Toronto Press}.
\newblock


\bibitem[Sonbol(2005a)]%
        {sonbol1993}
\bibfield{author}{\bibinfo{person}{Amira El-Azhary Sonbol}.}
  \bibinfo{year}{2005}\natexlab{a}.
\newblock \bibinfo{booktitle}{\emph{Beyond the Exotic: Women's Histories in
  Islamic Societies}}.
\newblock \bibinfo{publisher}{Syracuse University Press}, Chapter Introduction.
\newblock


\bibitem[Sonbol(2005b)]%
        {sonbol2005}
\bibfield{author}{\bibinfo{person}{Amira El~Azhary Sonbol}.}
  \bibinfo{year}{2005}\natexlab{b}.
\newblock \bibinfo{booktitle}{\emph{Beyond the Exotic: Women’s Histories in
  Islamic Societies}}.
\newblock \bibinfo{publisher}{Syracuse University Press},
  \bibinfo{address}{Thousand Oaks, CA}.
\newblock


\bibitem[Srinivasan and Uchino(2021)]%
        {Srinivasan2021-zd}
\bibfield{author}{\bibinfo{person}{Ramya Srinivasan} {and}
  \bibinfo{person}{Kanji Uchino}.} \bibinfo{year}{2021}\natexlab{}.
\newblock \showarticletitle{Biases in generative art: A causal look from the
  lens of art history}. In \bibinfo{booktitle}{\emph{Proceedings of the 2021
  {ACM} Conference on Fairness, Accountability, and Transparency}}.
  \bibinfo{pages}{41--51}.
\newblock


\bibitem[Suresh et~al\mbox{.}(2022)]%
        {suresh2022}
\bibfield{author}{\bibinfo{person}{Harini Suresh}, \bibinfo{person}{Rajiv
  Movva}, \bibinfo{person}{Amelia~Lee Dogan}, \bibinfo{person}{Rahul Bhargava},
  \bibinfo{person}{Isadora Cruxen}, \bibinfo{person}{Angeles~Martinez Cuba},
  \bibinfo{person}{Guilia Taurino}, \bibinfo{person}{Wonyoung So}, {and}
  \bibinfo{person}{Catherine D'Ignazio}.} \bibinfo{year}{2022}\natexlab{}.
\newblock \showarticletitle{Towards Intersectional Feminist and Participatory
  ML: A Case Study in Supporting Feminicide Counterdata Collection}. In
  \bibinfo{booktitle}{\emph{2022 ACM Conference on Fairness, Accountability,
  and Transparency}} (Seoul, Republic of Korea) \emph{(\bibinfo{series}{FAccT
  '22})}. \bibinfo{publisher}{Association for Computing Machinery},
  \bibinfo{address}{New York, NY, USA}, \bibinfo{pages}{667–678}.
\newblock
\showISBNx{9781450393522}
\urldef\tempurl%
\url{https://doi.org/10.1145/3531146.3533132}
\showDOI{\tempurl}


\bibitem[Timmermans and Tavory(2012)]%
        {timmermans2012theory}
\bibfield{author}{\bibinfo{person}{Stefan Timmermans} {and}
  \bibinfo{person}{Iddo Tavory}.} \bibinfo{year}{2012}\natexlab{}.
\newblock \showarticletitle{Theory construction in qualitative research: From
  grounded theory to abductive analysis}.
\newblock \bibinfo{journal}{\emph{Sociological theory}} \bibinfo{volume}{30},
  \bibinfo{number}{3} (\bibinfo{year}{2012}), \bibinfo{pages}{167--186}.
\newblock


\bibitem[Tomasev et~al\mbox{.}(2022)]%
        {tomasev2022manifestations}
\bibfield{author}{\bibinfo{person}{Nenad Tomasev},
  \bibinfo{person}{Jonathan~Leader Maynard}, {and} \bibinfo{person}{Iason
  Gabriel}.} \bibinfo{year}{2022}\natexlab{}.
\newblock \showarticletitle{Manifestations of Xenophobia in AI Systems}.
\newblock \bibinfo{journal}{\emph{arXiv preprint arXiv:2212.07877}}
  (\bibinfo{year}{2022}).
\newblock


\bibitem[Vaughn et~al\mbox{.}(2012)]%
        {Vaughn2012-ij}
\bibfield{author}{\bibinfo{person}{Sharon~R Vaughn},
  \bibinfo{person}{Jeanne~Shay Schumm}, {and} \bibinfo{person}{Jane~M
  Sinagub}.} \bibinfo{year}{2012}\natexlab{}.
\newblock \bibinfo{booktitle}{\emph{Focus group interviews in education and
  psychology}}.
\newblock \bibinfo{publisher}{SAGE Publications}, \bibinfo{address}{Thousand
  Oaks, CA}.
\newblock


\bibitem[Wang et~al\mbox{.}(2022)]%
        {wang2022}
\bibfield{author}{\bibinfo{person}{Angelina Wang}, \bibinfo{person}{Solon
  Barocas}, \bibinfo{person}{Kristen Laird}, {and} \bibinfo{person}{Hanna
  Wallach}.} \bibinfo{year}{2022}\natexlab{}.
\newblock \showarticletitle{Measuring Representational Harms in Image
  Captioning}. In \bibinfo{booktitle}{\emph{2022 ACM Conference on Fairness,
  Accountability, and Transparency}} (Seoul, Republic of Korea)
  \emph{(\bibinfo{series}{FAccT '22})}. \bibinfo{publisher}{Association for
  Computing Machinery}, \bibinfo{address}{New York, NY, USA},
  \bibinfo{pages}{324–335}.
\newblock
\showISBNx{9781450393522}
\urldef\tempurl%
\url{https://doi.org/10.1145/3531146.3533099}
\showDOI{\tempurl}


\bibitem[Weidinger et~al\mbox{.}(2022)]%
        {Weidinger2022}
\bibfield{author}{\bibinfo{person}{Laura Weidinger}, \bibinfo{person}{Jonathan
  Uesato}, \bibinfo{person}{Maribeth Rauh}, \bibinfo{person}{Conor Griffin},
  \bibinfo{person}{Po-Sen Huang}, \bibinfo{person}{John Mellor},
  \bibinfo{person}{Amelia Glaese}, \bibinfo{person}{Myra Cheng},
  \bibinfo{person}{Borja Balle}, \bibinfo{person}{Atoosa Kasirzadeh},
  \bibinfo{person}{Courtney Biles}, \bibinfo{person}{Sasha Brown},
  \bibinfo{person}{Zac Kenton}, \bibinfo{person}{Will Hawkins},
  \bibinfo{person}{Tom Stepleton}, \bibinfo{person}{Abeba Birhane},
  \bibinfo{person}{Lisa~Anne Hendricks}, \bibinfo{person}{Laura Rimell},
  \bibinfo{person}{William Isaac}, \bibinfo{person}{Julia Haas},
  \bibinfo{person}{Sean Legassick}, \bibinfo{person}{Geoffrey Irving}, {and}
  \bibinfo{person}{Iason Gabriel}.} \bibinfo{year}{2022}\natexlab{}.
\newblock \showarticletitle{Taxonomy of Risks Posed by Language Models}. In
  \bibinfo{booktitle}{\emph{2022 ACM Conference on Fairness, Accountability,
  and Transparency}} (Seoul, Republic of Korea) \emph{(\bibinfo{series}{FAccT
  '22})}. \bibinfo{publisher}{Association for Computing Machinery},
  \bibinfo{address}{New York, NY, USA}, \bibinfo{pages}{214–229}.
\newblock
\showISBNx{9781450393522}
\urldef\tempurl%
\url{https://doi.org/10.1145/3531146.3533088}
\showDOI{\tempurl}


\bibitem[Weinberg(2022)]%
        {weinberg2022rethinking}
\bibfield{author}{\bibinfo{person}{Lindsay Weinberg}.}
  \bibinfo{year}{2022}\natexlab{}.
\newblock \showarticletitle{Rethinking Fairness: An Interdisciplinary Survey of
  Critiques of Hegemonic ML Fairness Approaches}.
\newblock \bibinfo{journal}{\emph{Journal of Artificial Intelligence Research}}
   \bibinfo{volume}{74} (\bibinfo{year}{2022}), \bibinfo{pages}{75--109}.
\newblock


\bibitem[Wolfe et~al\mbox{.}(2022)]%
        {wolfe2022contrastive}
\bibfield{author}{\bibinfo{person}{Robert Wolfe}, \bibinfo{person}{Yiwei Yang},
  \bibinfo{person}{Bill Howe}, {and} \bibinfo{person}{Aylin Caliskan}.}
  \bibinfo{year}{2022}\natexlab{}.
\newblock \showarticletitle{Contrastive Language-Vision AI Models Pretrained on
  Web-Scraped Multimodal Data Exhibit Sexual Objectification Bias}.
\newblock \bibinfo{journal}{\emph{arXiv preprint arXiv:2212.11261}}
  (\bibinfo{year}{2022}).
\newblock


\bibitem[Yee et~al\mbox{.}(2021)]%
        {Yee2021-wr}
\bibfield{author}{\bibinfo{person}{Kyra Yee}, \bibinfo{person}{Uthaipon
  Tantipongpipat}, {and} \bibinfo{person}{Shubhanshu Mishra}.}
  \bibinfo{year}{2021}\natexlab{}.
\newblock \showarticletitle{Image Cropping on Twitter: Fairness Metrics, their
  Limitations, and the Importance of Representation, Design, and Agency}.
\newblock \bibinfo{journal}{\emph{Proc. ACM Hum.-Comput. Interact.}}
  \bibinfo{volume}{5}, \bibinfo{number}{CSCW2} (\bibinfo{date}{Oct.}
  \bibinfo{year}{2021}), \bibinfo{pages}{1--24}.
\newblock


\bibitem[Yu et~al\mbox{.}(2022a)]%
        {yu2022scaling}
\bibfield{author}{\bibinfo{person}{Jiahui Yu}, \bibinfo{person}{Yuanzhong Xu},
  \bibinfo{person}{Jing~Yu Koh}, \bibinfo{person}{Thang Luong},
  \bibinfo{person}{Gunjan Baid}, \bibinfo{person}{Zirui Wang},
  \bibinfo{person}{Vijay Vasudevan}, \bibinfo{person}{Alexander Ku},
  \bibinfo{person}{Yinfei Yang}, \bibinfo{person}{Burcu~Karagol Ayan},
  {et~al\mbox{.}}} \bibinfo{year}{2022}\natexlab{a}.
\newblock \showarticletitle{Scaling autoregressive models for content-rich
  text-to-image generation}.
\newblock \bibinfo{journal}{\emph{arXiv preprint arXiv:2206.10789}}
  (\bibinfo{year}{2022}).
\newblock


\bibitem[Yu et~al\mbox{.}(2022b)]%
        {parti}
\bibfield{author}{\bibinfo{person}{Jiahui Yu}, \bibinfo{person}{Yuanzhong Xu},
  \bibinfo{person}{Jing~Yu Koh}, \bibinfo{person}{Thang Luong},
  \bibinfo{person}{Gunjan Baid}, \bibinfo{person}{Zirui Wang},
  \bibinfo{person}{Vijay Vasudevan}, \bibinfo{person}{Alexander Ku},
  \bibinfo{person}{Yinfei Yang}, \bibinfo{person}{Burcu~Karagol Ayan},
  \bibinfo{person}{Ben Hutchinson}, \bibinfo{person}{Wei Han},
  \bibinfo{person}{Zarana Parekh}, \bibinfo{person}{Xin Li},
  \bibinfo{person}{Han Zhang}, \bibinfo{person}{Jason Baldridge}, {and}
  \bibinfo{person}{Yonghui Wu}.} \bibinfo{year}{2022}\natexlab{b}.
\newblock \bibinfo{title}{Scaling Autoregressive Models for Content-Rich
  Text-to-Image Generation}.
\newblock
\newblock
\urldef\tempurl%
\url{https://doi.org/10.48550/ARXIV.2206.10789}
\showDOI{\tempurl}


\end{thebibliography}
\newpage
\begin{appendices}
\section{Supplemental Methods}
\label{sec:supplementary_methods}
\subsection{Operationalizing Culture}

Since our recruitment and procedures relied on  an operationalization of the term culture, we share our definitions and scoping rationales. “Culture” is often a stand in for many types of ideas and objects. Sociologically, it usually refers to: norms, values, beliefs, expressive symbols, or practices (behavior patterns) \cite{Griswold2013}. Culture is dynamic and multiple, and is interwoven into political and economic aspects of the social world \cite{Murphie2017}.  Our study necessitated two operationalizations of culture: one for our recruitment (\textit{what is a cultural group?}) and second for our prompt engineering (\textit{what is a cultural category?}). 

The cultural group or community definition we chose relied on geographic and ethnic bounds.  Ethnicity has to do with the politicization of culture, involving shared ``cultural heritage, ancestry, origin myth, homeland, language, dialect, symbolic systems.’’ These often map on to nation states, geographies, and languages. Given we wanted to localize our understanding of culture, we localized our recruitment criteria. Instead of treating South Asia as a homogenous group, we divided the context by nation states and recruited from three different South Asian contexts: Bangladesh, India and Pakistan. We chose to then conduct the focus groups stratified along these nation state lines understanding the historic and political tension between the three states.  Their geographical proximity and shared borders means cultural practices and languages bleed into each other. Historically shifting borders and empires meant these countries were often part of the same political aggregations. While these three countries emerge from similar historic conditions and have overlap in cultures over time, they have also developed their own cultures and internal understanding of them due to stratified infrastructure, (e.g., education systems). For example, it was unlikely that people in the Indian focus groups would have had knowledge of cultural norms and traditions in common with regional groups in Pakistan or Bangladesh. We wanted to have participants from each nation state  focus on the histories and cultures most relevant to them, with enough common overlap to facilitate a cohesive yet nuanced analysis.  However, we did not require them to be citizens or residents of any of the nation states. This allowed us to capture participants with deep knowledge of the context while also including the South Asian diaspora who may no longer be living in these countries. 

To define cultural categories for prompt engineering, we relied on existing empirical studies which offered no set definition of culture, but which broadly divided culture into capital-C and small-c. The former comprises of material aspects of cultural production (e.g., TV/film,architecture). Small-c culture is less tangible by referring to shared norms, values, and history. Thus we intentionally incorporated elements of both types of culture into the prompts we tested during this study. You can see a table below of the different elements of culture we showed participants in FG1 to ground our conversations  
\subsection{Methods and Procedures}
We aimed to keep the focus group environment safe while encouraging challenging discussions in line with our community-centered approach. As such we made some intentional choices while designing the focus groups. First, we were mindful to keep conversations at a size that would allow everyone to contribute. When focus groups had more than six participants, we held large group discussions at the beginning and end, but opted to split the group into two breakout rooms during the key activities when participant contributions were sought. Additionally, we offered multiple formats to receive contributions including: verbal, video conference chat, written feedback on slides, anonymous written feedback on a virtual whiteboard, and through the interim survey. We also developed communication considerations to remind participants that harassment would not be tolerated, encourage participants to be mindful of their contributions, and to open opportunities for others. The lead facilitator was mindful to make time for everyone to contribute if they wished, and to respond to comments made via the various contribution formats. Within each focus group 1-3 additional researchers were present to troubleshoot technical issues and take notes, but remained quiet to further center participant perspectives.
The following subsections detail the procedures during each focus group and the interim survey.
\subsubsection{Focus Group I}
The first focus group opened with a high-level introduction to text-to-image technologies and an overview of the study goals. Participants also introduced themselves and the focus group moderators outlined guidelines for inclusive and respectful dialogue. The remainder of the session was structured around discussion questions and interactive virtual whiteboard activities to learn how participants characterized ‘good’ and ‘bad’ cultural representations they had encountered in media and technology, and how participants might assess ‘good’ or ‘bad’ representations in AI-generated imagery. 
\subsubsection{Interim Survey}
Between the first and second focus groups, participants completed a survey. The survey asked for five specific prompts (e.g., ``People playing cricket in streets of Balochistan’’) and an explanation of why the prompt, and which cultural elements it referenced, that were important to them and what whey hoped to learn about T2I models by testing the prompt. 
Participants were also asked to share up to 5 specific elements of their cultures that could be incorporated into text prompts to enable evaluation of the models’ ability to produce respectful cultural representations. These elements fit into the following categories: 
\begin{itemize}
   \item Cultural events/holidays/festivals/rituals 
   \item Landmarks
   \item Art styles/artists
   \item Historic events/figures
   \item Characters or stories from fiction/folklore/literature/film
\end{itemize}
We also utilized the survey as an opportunity to solicit feedback about participant experiences in the first focus group, and any ways we could improve their experiences in the second focus group. 

\subsubsection{Focus Group II}
Held three weeks after Focus Group I., participants reflected on specific T2I images generated from prompts that the research team constructed based on their suggestions posed during the first focus group and the survey. Specifically, they spent individual time providing written feedback on images rendered from prompts based on their personal suggestions. Feedback related to how well the generated images represented the prompt and how well or not they aligned with participant expectations based on their own cultural knowledge. We then reconvened the group for facilitated discussion, where participants others present could expand their own, or others, feedback and discuss higher level excitements and concerns around representations and responsible T2I development.
\subsubsection{Prompt Engineering}
Table \ref{tab:prompts} shows examples of participant-suggested prompts. We developed a set of prompts to generate images for the second focus group by synthesizing participant-suggested prompts. We sought to increase the quality and coverage of cultural references in the study. We tested prompts to avoid including those that resulted in non-cultural failures since we were interested in exploring cultural limitations specifically rather than broad limitations of T2I models. Testing the prompts ahead of time helped ensure we made best use of participant’s time and expertise. In some cases, we synthesized suggested prompts from participants into a single prompt to increase our coverage of cultural references. For instance, if a participant suggested a prompt containing only one artist or art style, and then mentioned a type of clothing as a second prompt, we sometimes synthesized the two into a single prompt of “clothing artifact painted in the style of artist.” However, we also included many prompts that referenced a single cultural artifact, given that merging references can pose challenges to identifying the ``cause’’ of a failure, e.g.  “\texttt{[clothing artifact]} painted in the \texttt{[style of artist]}” may fail to generate both the artifact \textit{and} the style, due to a failure to recognize the art style, whereas “\texttt{[clothing artifact]}” on it’s own may produce an appropriate rending of the artifact.   

Notably, we did not engineer prompts to ensure systematic coverage over different cultural categories, nor did we systematically disentangle different cultural references within the prompts. We leave this important line of inquiry to future work that can offer deeper insights into which cultural elements cause failures and the prevalence of such failures across models. 

\begin{table}[t]
\centering
    \begin{tabular}{l p{13pc}}
    \hline
    \multirow{4}{*}{India} &
    A Kalamkari saree drying in the sun in a terrace garden \\
    & Children eating fried street food in Varanasi \\
    & An Indian wedding \\
    & An affluent family in South Asian family\\
    \hline
    \multirow{4}{*}{Pakistan} &
    A scene from daily life in Karachi \\
    & A scene from village life in Punjab in the spring \\
    & A genius at work \\
    & Truck art \\
    \hline
    \multirow{4}{*}{Bangladesh} &
    Deshi people celebrating Pohela Boisakh by the Jaflong Sea \\
    & Students enjoying tea and snacks at a Dhaka University open air coffee shop \\
    & Modern women wearing garmeen check sharis at a fish market \\
    & Deshi people \\
    \hline
    \end{tabular}
    \caption{Example prompts used to generate images for focus groups, broken down by country. Prompts were constructed based on participant suggestions, with minor wording changes and occasional merging of concepts. Notably, some prompts omit any references to cultural groups (e.g. ``A genius at work'') whereas others include references to high level cultural groups (e.g. ``An affluent family in \textit{South Asian} family'') of more granular cultural groups (e.g. ``\textit{Deshi people} celebrating Pohela Boisakh by the Jaflong Sea''). Such granularization of cultural references enables us to explore the hierarchy of cultural defaults reflected in T2I images. }
    \label{tab:prompts}
\end{table}
\subsection{Participant Demographics}
Thirty-six participants took part in our study (Pakistan (\textit{n} = 15); India (\textit{n} = 13); Bangladesh (\textit{n} = 8). Table \ref{tab:gender} shows the gender breakdown per country, Table \ref{tab:age} shows the age breakdown per country, and Table \ref{tab:occupation} shows the occupation category breakdown by country.

\begin{table}
\vspace{2mm}
    \small
    \centering
    \begin{tabular}{c|c|c|c}
     &
    India &
    Pakistan &
    Bangladesh \\
    \hline\hline
    Male &
    5 &
    5 &
    2 \\
    \hline
    Female  &
    7 &
    8 &
    4 \\
    \hline
    Not specified  &
    1 &
    2 &
    2 \\
    \hline
    \end{tabular}
    \caption{Study participants by self-identified gender and country.}
    \label{tab:gender}
\end{table}

\begin{table}
    \small
    \centering
    \begin{tabular} {c|c|c|c}
     &
    India &
    Pakistan &
    Bangladesh \\
    \hline\hline
    18-24 &
    0 &
    1 &
    0 \\
    \hline
    25-34  &
    8 &
    8 &
    4 \\
    \hline
    35-44  &
    4 &
    3 &
    2 \\
    \hline
    45-54 &
    1 &
    3 &
    2 \\
    \hline
    \end{tabular}
    \caption{Study participants by age and country.}
    \label{tab:age}
\end{table}

\begin{table}
\small
    \begin{tabular}{c|c|c|c}
     &
    India &
    Pakistan &
    Bangladesh \\
    \hline\hline
    Academic researchers &
    9 &
    7 &
    10 \\
    \hline
    Cultural workers (Non-academic)  &
    4 &
    2 &
    3 \\
    \hline
    Non-cultural workers  &
    2 &
    4 &
    4 \\
    \hline
    \end{tabular}
    \caption{Study participants by occupation.}
    \label{tab:occupation}
\end{table}



\section{Supplemental Images}
\label{sec:appendix_images}
\begin{figure}
    \begin{minipage}{.4\textwidth}
        \centering
        \begin{subfigure}[t!]{0.4\textwidth}
            \centering
            \hspace{10mm}
            \includegraphics[height=0.95in]{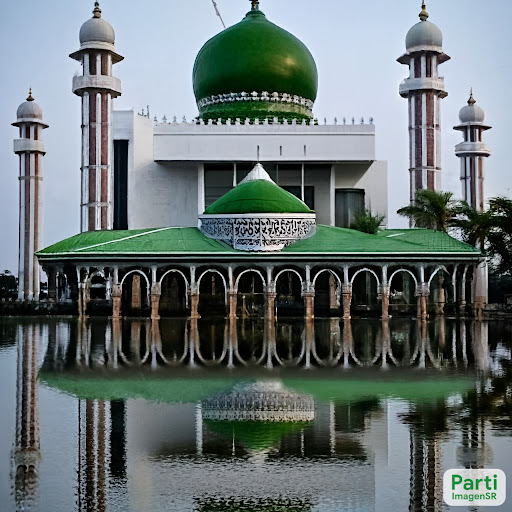}
            \caption{\footnotesize{}}
        \end{subfigure}%
        \begin{subfigure}[t!]{0.4\textwidth}
            \centering
            \includegraphics[height=0.95in]{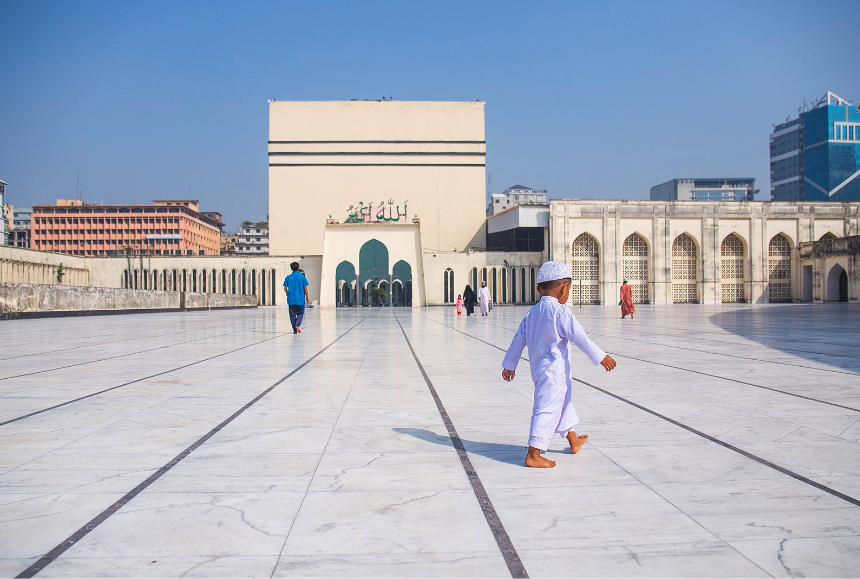}
            \captionsetup{justification=centering}
            \caption{\footnotesize{}}
        \end{subfigure}
        \captionsetup{justification=justified, margin=0cm}
        \caption{Example of (a) a Parti generated image for the prompt ``Baitul Mukarram National Masjid'' juxtaposed with (b)  a photograph of Baitul Mukarram National Masjid \cite{Baitul}. The generated image lacks the distinct architectural features of Baitul Mukarram National Masjid.}
        \Description [Two images, one T2I-generated and the other canonical,  each of Baitul Mukarram National Masjid.]{The T2I image depicts a green and white mosque with a reflective pool in the foreground. A large green dome is centered, and the walls feature ornamental arches  Two minarets with smaller green domes flank each side. The Parti watermark  is written at the bottom right.

The canonical image is a low perspective of a long, white marble pathway leading to a courtyard where people pass through and a white, square mosque in the background. It has an arched  glass door, and the walls feature  horseshoe arches.   }
        \label{fig:failure_to_rec_appendix-1}
    \end{minipage}
\end{figure}

\begin{figure}
    \begin{minipage}{.4\textwidth}
        \centering
        \begin{subfigure}[t!]{0.45\textwidth}
            \centering
            \includegraphics[height=0.95in]{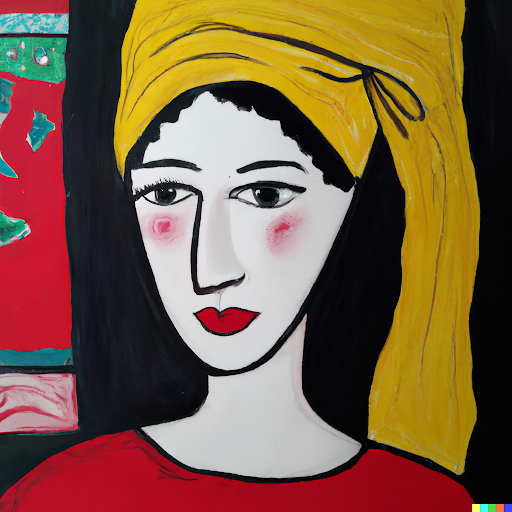}
            \captionsetup{justification=centering,margin=2cm}
            \caption{\footnotesize{}}
        \end{subfigure}%
        \begin{subfigure}[t!]{0.4\textwidth}
            \centering
            \includegraphics[height=0.95in]{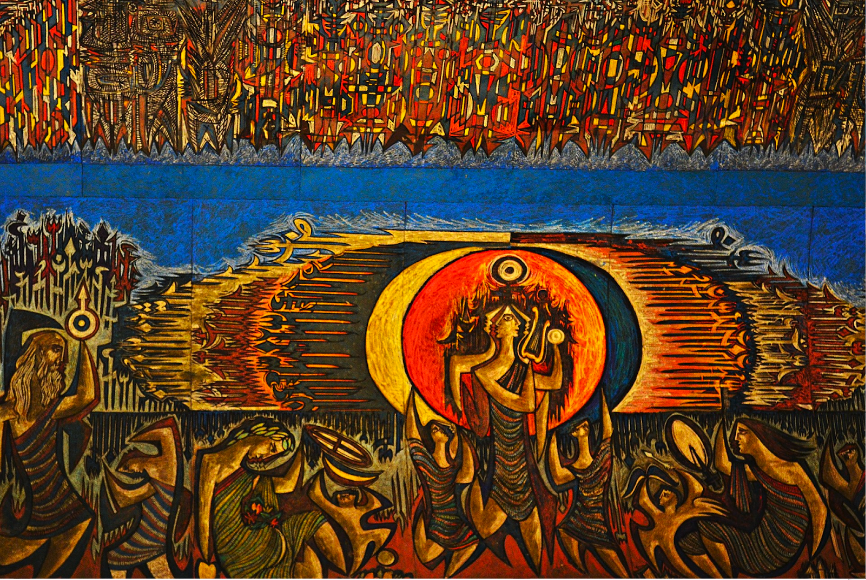}
            \captionsetup{justification=centering,margin=2cm}
            \caption{\footnotesize{}}
        \end{subfigure}
        \caption{Example of (a) a DALL-E generated image for the prompt ``A woman in the style of artist Sadequain'' juxtaposed with (b) a painting by Sadequain \cite{Sadequain}. The generated image does not capture the artist's style, notably lacking the distinctive calligraphy and color themes.}
        \Description [Two images, one T2I-generated in the style of Sadequain, and the other a canonical mural by Sadequain.]{The T2I image depicts a woman in portrait. She is wearing a red top. Her black curly hair peeks out of a gold head scarf that drapes down her right side. She has a neutral look on her face, gazing off to the left. She wears red lipstick and has rosy cheeks, long gray hair, and a yellow face. Her eyes are large and close together. Below her eyes is a white diamond. Within the diamond is a red rectangle in place of a nose, and the top of one eyelid comes down into the white diamond. A red circle serves as her mouth. The left half of her thick neck is black; the right half is red. The same black and red make up the background. DALL-E watermark is near the bottom right.

The mural is shown on the  Ceiling of Frere Hall. The bottom third features a woman wearing muted blue and burnt orange garb, depicted in 3 overlapping layers which give the illusion that the woman’s face moves from left to right. Each layer features a face with 1, centered  eye. The face then splits as if it were looking in both directions; each side has its own nose and mouth. The woman in the back most layer appears to have 3 arms; one holds up a harp; a golden disk in another. The arms are bent differently in the other two layers to fill remaining space. Above her head is a small circle. A large orange circle is behind her, and extends into intricate, abstract shapes that continue up the top two thirds of the image, featuring colors including sea blue, brown and gold. Nine dancers of different ages from children to elder, some holding instruments, extend across the image’s bottom. An elder man holds up the ‘male’ symbol.}
        \label{fig:failure_to_rec_appendix-2}
    \end{minipage}
\end{figure}

Here, we present additional examples to showcase the participant identified model limitations discussed in Section \ref{sec:limitations}. Figures \ref{fig:failure_to_rec_appendix-1} and \ref{fig:failure_to_rec_appendix-2} show examples of \textit{failing to recognize cultural subjects}. Figure \ref{fig:failure_to_rec_appendix-2} shows how the model generated a generic green domed mosque for the prompt ``Baitul Mukarram National Masjid'', which we have contrasted with a photograph of the real Baitul Mukarram National Mosque in Dhaka which does not have any of the features the model generated. Similarly, Figure \ref{fig:failure_to_rec_appendix-2} shows a comparison between one work of a  famous Pakistani artist Sadequain and the model's rendition of  Sadequain's work. The model failed to recognize Sadequain as indicated by absence of any stylistic features common to Sadequain's work. 

\enlargethispage{30pt}

\begin{figure}
    \centering
        \includegraphics[width=.45\textwidth]{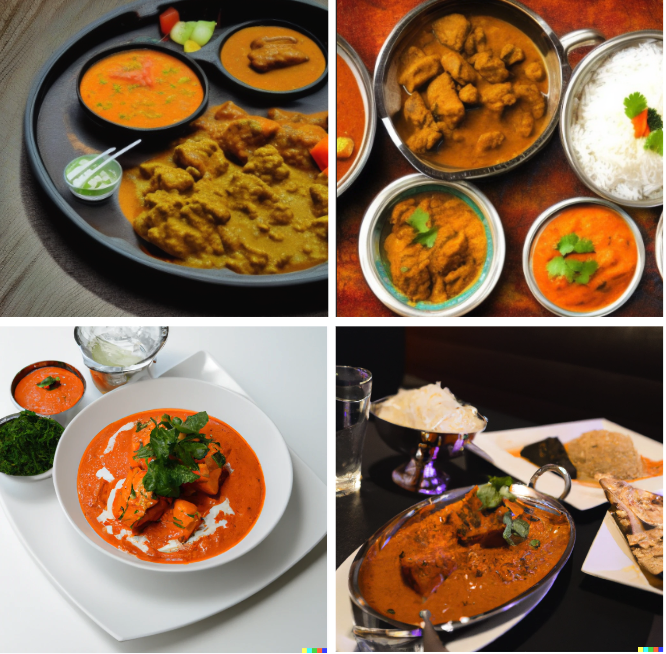}
         \caption{Generated images, from Stable Diffusion and DALL-E, for the prompt "Indian Food" showing exclusively North Indian cuisine and failing to depict the diversity of food.}
         \Description[T2I-generated images of Indian food.]{Top left: A closeup of a black, circular plate on a dark wooden table. On the plate is a large portion of meat curry, in an orangish-brown sauce. Two small black bowls are arranged above the curry, on the plate, with a bright orange curry topped with green herbs in the left bowl and a dark orange curry in the right bowl. A small bowl with green herb garnish is on the plate’s far left.

Top right: An overhead shot of a red table, with five family-style bowls. The largest bowl has meat in a brown curry sauce. To the right is a large bowl of steamed rice with cilantro garnish, and the smaller bowls below contain what appears to be different types of meat curry, garnished with green herbs.

Bottom left: A closeup of a black table with a large bowl of orange curry garnished with green herbs, with a plate of naan beside. Behind is a heaping silver bowl of rice, and a square plate with a biryani-resembling rice dish is further back. A tall glass of water is on the table’s left side. DALL-E watermark is near the bottom right.

Bottom right: A tikka masala-resembling dish topped with cilantro is in a white bowl, sitting on a white plate, placed on top of a white tablecloth. Beside the plate are three small side bowls resembling a mint chutney, a bright red/orange sauce, and a white sauce resembling raita. DALL-E watermark is near the bottom right.}
         \label{fig:defaults-1}
\end{figure}

\begin{figure}
        \centering
        \includegraphics[height=2in]{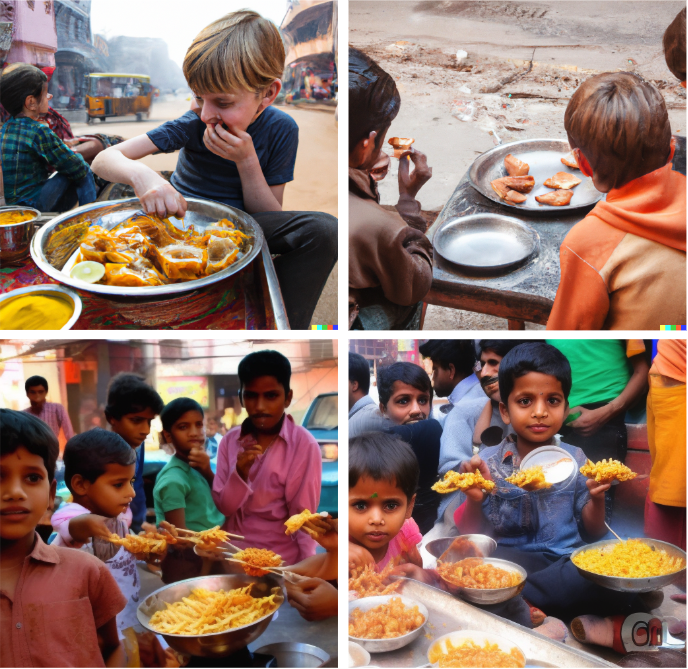}
         \caption{Generated images,from Stable Diffusion and DALL-E, for the prompt "Children eating fried street food in Varanasi" showing multiple light-skinned children.}
         \Description [T2I -generated images of children eating street food in Varanasi.]{Upper left: A closeup of a young boy with white skin and short blonde hair and blue t-shirt, who shovels nondescript food out of a large bowl into his mouth. A dusty city street with a rickshaw is in the background. DALL-E watermark is near the bottom right.

Upper right: The back head and upper bodies of a young dark-skinned and dark-haired child (left) and a young white-skinned and blond-haired child (right). A plate of nondescript food is in front of them.  DALL-E watermark is near the bottom right.

Lower left: A group of young dark-skinned and dark-haired  boys eating nondescript food from a large bowl. 

Lower right: A group of young dark-skinned and dark-haired  boys eating nondescript food from a large bowl, with several other children holding food in their hands.}
         \label{fig:defaults-2}
    \end{figure}

Figures \ref{fig:defaults-1} and \ref{fig:defaults-2} illustrate \textit{cultural defaults}. Figure \ref{fig:defaults-1} showcases North Indian as a cultural default, where prompts for ``Indian food'' returned only food participants associated with North India. Figure \ref{fig:defaults-2} showcases a Western default, where even for street images in Varanasi, a town in India, the model inserted white skinned children. 

\begin{figure}
    \centering
        \includegraphics[height=2in]{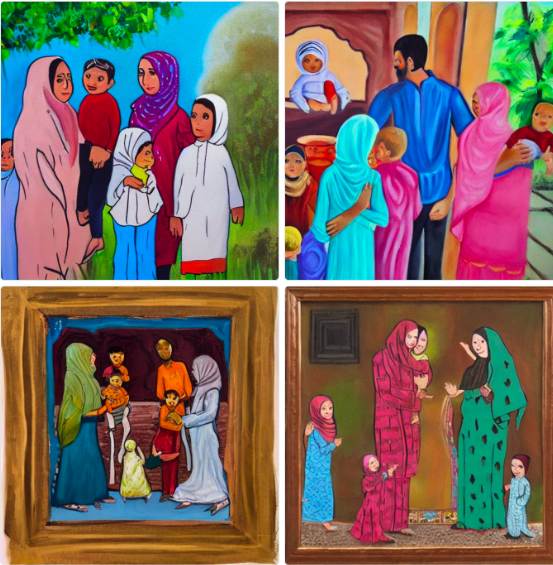}
         \caption{Generated images, from Stable Diffusion, for the prompt "A Family Celebrating Eid" only showing women, even children, in headscarves and religiously traditional clothes, reflecting the trope of Muslim lives as one-dimensional. }
         \Description [T2I-generated images of a family celebrating Eid.]{Top left: An oil-style painting of two women and three young children. Two young girls are standing in front of the woman on the right. A young boy is held in the arms of the woman on the left. The women and young girls are all wearing headscarves.

Top right: An oil-style  painting of a family with a man, woman holding an infant, and six children, one partially obscured. The woman and girl children wear headscarves. The family embraces; the man’s arms are around the woman and a child, and two children hold each other. A small child sits on a tabletop, the man looking back at them. Plants and an outdoor patio are in the background. The people and embraces look somewhat disformed, an aberration of realistic T2I depictions of people.

Bottom left: A wooden-framed oil-styled painting of a family with two women,a man, and four young children. The women are each standing on opposite sides of the painting, facing inwards, with the others in between. The women and the young girl are wearing head scarves.

Bottom right: A wooden-framed oil-styled painting with two women centered next to three children, all close-mouthed smiling. The woman on the left wears a headscarf and traditional clothing with a red pattern and holds a child. She is in conversation with the woman on the right who also wears a headscarf and traditional, teal clothing. Two wear headscarves. The third child holds onto the right woman’s teal clothing. A brown wall with a decorative object with concentric squares is visible. The people look somewhat disformed, an aberration of realistic T2I depictions of people.}
         \label{fig:tropes-1}
    \end{figure}

\begin{figure}
        \centering
        \includegraphics[height=2in]{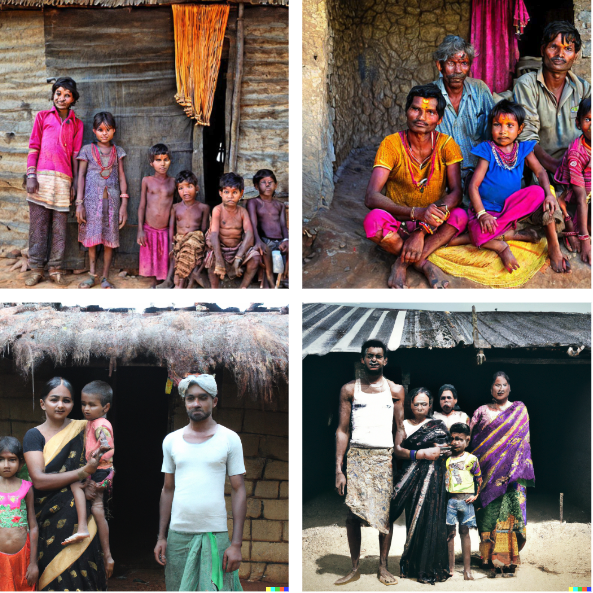}
         \caption{Generated images, from Stable Diffusion and DALL-E, for the prompt “A photo of a family in an Adivasi village” showing poor families and unkempt houses and reflecting the trope of South Asian as impoverished and under-developed. }
         \Description [T2I-generated images of a family in an Adivasi village.]{Top left: Six stoic children wearing South Asian-appearing clothing who are presented in order of height. They stand or sit in front of a run-down mud shack; fabric covers the opening. The children appear thin and to be wearing simple clothing. Three children have no shirts. The children look somewhat disformed, an aberration of realistic T2I depictions of people.

Top right: A stoic family of 2 children, 2 adult men, and one elder man wearing South Asian-appearing clothing, sitting on a muddy floor in the foreground in front of a mud and stone home. Everyone’s clothes, hands, and feet look dirty. The people look somewhat disformed, an aberration of realistic T2I depictions of people.

Bottom left: A stoic family of a man, a woman carrying a small child in her arms, and a young girl, all standing in front of a stone home. DALL-E watermark is near the bottom right.

Bottom right: Four adults and a small child standing outside a home with a metal roof. DALL-E watermark is near the bottom right.}
         \label{fig:tropes-2}
\end{figure}

Figures \ref{fig:tropes-1} and \ref{fig:tropes-2} illustrate \textit{cultural tropes}. Figure \ref{fig:tropes-1} shows how Muslim lives are reduced to one dimensionality of a particular religious stereotype where every woman is veiled and no richness or diversity of cultures is shown. Figure \ref{fig:tropes-2} shows generated images which reproduce a media trope of India's indigenous populations as dirty, unkempt and poor—almost as if they were images for a government ad campaign said one participant.  

       
\clearpage
\end{appendices}

\end{document}